\documentclass[11pt,a4paper]{article}
\pdfoutput=1
\usepackage{jcappub}
\usepackage{graphicx}
\usepackage{cprotect}
\usepackage{epstopdf}

\newcommand{\PRE}[1]{{#1}} 

\newcommand{\be}{\begin{equation}}
\newcommand{\ee}{\end{equation}}
\newcommand{\bea}{\begin{eqnarray}}
\newcommand{\eea}{\end{eqnarray}}

\def\gev{\text{ GeV}}

\def\kT{\text{ kT}}

\def\cm{\text{ cm}}
\def\m{\text{ m}}
\def\km{\text{ km}}

\def\s{\text{ s}}
\def\yr{\text{ yr}}

\def\beq{\begin{eqnarray}}
\def\eeq{\end{eqnarray}}
\def\bea{\begin{eqnarray}}
\def\eea{\end{eqnarray}}

\def\sigmaSD{\sigma_{\rm SD}}

\newcommand{\gsim}{\lower.7ex\hbox{$\;\stackrel{\textstyle>}{\sim}\;$}}
\newcommand{\lsim}{\lower.7ex\hbox{$\;\stackrel{\textstyle<}{\sim}\;$}}

\usepackage{enumerate}
\include{epsf}

\title{
\textsc{Neutrino Topology Reconstruction at DUNE and Applications to Searches for Dark Matter Annihilation in the Sun}
\PRE{\vspace*{0.1in}}
}

\author[a]{Carsten Rott}
\author[a]{DongYoung Jeong}
\author[b]{Jason Kumar}
\author[c]{David Yaylali}

\affiliation[a]{\mbox{Department of Physics, Sungkyunkwan University,} \\ \mbox{2066 Seobu-ro, Suwon 16419, Korea}}
\affiliation[b]{\mbox{Department of Physics \& Astronomy, University of
Hawai'i,} \\ \mbox{2505 Correa Road, Honolulu, HI 96822, U.S.A.}}
\affiliation[c]{\mbox{Department of Physics, University of Arizona,} \\ \mbox{1118 E. Fourth Street, Tucson, AZ 85721, U.S.A.}
\PRE{\vspace*{.1in}}
}

\emailAdd{rott@skku.edu}
\emailAdd{dongyoungjeong@gmail.com}
\emailAdd{jkumar@hawaii.edu}
\emailAdd{yaylali@email.arizona.edu}

\abstract{
We consider a new technique for neutrino energy and topology reconstruction at DUNE.  In particular, we show that
when the direction of the incoming neutrino is known, one can use the measured directions of the outgoing
leptonic and hadronic particles to reconstruct poorly-measured quantities, such as the hadronic cascade
energy.  We show that this alternative technique yields an energy resolution which is comparable to current reconstruction methods which sum measured energies. As a proof of concept we apply this new reconstruction method to a search for dark matter annihilation in the Sun. We show that the use of directional information from both the leptonic and hadronic interaction products allows one to effectively reject backgrounds and isolate the signal, giving competitive sensitivities.
}


\begin{document}

\begin{flushright}
{\large \tt
UH-511-1300-18}
\end{flushright}

\maketitle
\flushbottom
\date{\today}


\section{Introduction}

An important challenge in the data analysis of neutrino detectors is the neutrino energy reconstruction
 from the interaction products produced when a neutrino interacts in
the detector. The reconstruction efficiency and corresponding energy resolution have a critical
 impact on the detector's science capabilities.
As an example, neutrino detectors play an essential role in the search for particle dark matter~\cite{Bertone:2004pz}. In particular the search
for neutrinos which can arise if dark matter annihilates in the center of the Sun~\cite{Silk:1985ax,Press:1985ug,Krauss:1985ks} has been extremely competitive in constraining dark matter particle properties~\cite{Aartsen:2016exj,Aartsen:2017ulx,In:2017kcf,Choi:2015ara,Adrian-Martinez:2016gti,Avrorin:2014swy}.
If neutrino energies can be determined precisely, the observed energy spectrum will contain clues about the properties of the dark matter particle, such as its mass and annihilation
channels. Neutrino energy reconstruction and topology reconstruction are
also essential to discriminate a signal from the background arising from atmospheric neutrinos.

Liquid argon time-projection chamber (LArTPC) neutrino detectors, such as DUNE~\cite{Acciarri:2016ooe,Acciarri:2015uup}, offer  unprecedented opportunities through their excellent capabilities of individual particle tracking for event reconstructions. Previous sensitivity studies often relied for simplicity on reconstructing the energy of a neutrino by summing up all of the visible energy deposited in the detector.  In this work, we exploit the good particle identification along with the excellent angular and energy reconstruction properties of a LArTPC, which have already been proven by ICARUS~\cite{Amerio:2004ze} and ArgoNeuT~\cite{Anderson:2012vc} detectors.
We point out that,
if the direction of an incoming neutrino is known and if the directions of the visible particles
can be measured, then one can use kinematic constraints to improve the accuracy of energy
reconstruction algorithms. In particular, one can use energy and momentum conservation
to estimate the energy missed in the hadronic cascades produced by a deep-inelastic scatter.
Not only does this algorithm improve the precision of neutrino energy reconstruction, but also
significantly reduces the atmospheric neutrino background by allowing one to reject events whose
kinematics are inconsistent with neutrinos arriving from the direction of the source.


Neutrino event reconstruction is critical to measure neutrino oscillations, neutrino cross sections, or search for astrophysical neutrinos. A variety of event reconstruction methods have been developed for various energy ranges, that rely on the measured charge and hit times at the photo sensors of the detector~\cite{Ankowski:2015jya, Aartsen:2013vja, Missert:2017qdz}. For a given event topology hypothesis it is possible to produce a charge and time PDF for each sensor module. Different event hypotheses can be distinguished based on the observed data by comparing best-fit likelihoods. Deep inelastic scattering neutrino events are often reconstructed using the total deposited energy~\cite{Mousseau:2016snl}. Kinematic constraints imposed by an assumed two-body scattering process to extract the neutrino energy from the lepton energy and angle can be utilized in neutrino oscillation experiments~\cite{Abe:2017vif}. Transverse kinematics have been exploited to investigate nuclear effects in neutrino-nucleus scattering~\cite{Abe:2018pwo,Dolan:2018zye}.  A detector like DUNE will in many ways be very different from previous detectors: it will be capable of fine-grained tracking of large numbers of particle; the beam direction will be known, constraining the incoming neutrino direction and constraining its energy, to some extent. The event reconstruction challenge faced by DUNE is expected to resemble in many ways that of tracking chambers of collider physics experiments.


The use of such kinematic constraints in energy reconstruction is familiar in the context
of collider physics.  A notable example is the use of the ditau mass variable~\cite{Ellis:1987xu} to reconstruct
the mass of a parent particle (such as a $Z$-boson) which decays to a boosted $\tau^{+} \tau^{-}$ pair, with
each $\tau$ decaying leptonically.  In this case, the energy of the parent cannot be directly measured
by summing the energy of its decay products, since $\tau$ decay produces neutrinos which are not
measured.  Instead, the key to the energy reconstruction analysis is to note that one can use
momentum conservation in the plane transverse to the beam to solve for two unknowns.  If one assumes that all of
the missing energy arises from neutrinos which are collinear with either one of the boosted charged leptons
produced by decay of the $\tau^{+}$ and $\tau^{-}$, then there are only two unknowns, namely, the
energy of the missed neutrinos collinear with each lepton.  The ditau mass variable is the solution for
the squared mass of the parent in terms of well-measured observables, under the assumption of this particular
decay topology, and can be used not only to reconstruct the mass of the parent, but also to reject
events in which the underlying process has a different topology.

We will be able to reconstruct the neutrino energy in a similar way.  In particular, we will
assume that a neutrino with an unknown energy, but fixed direction, has a deep-inelastic scatter
against a parton with unknown momentum fraction in the nucleus, producing a charged lepton with
well-measured energy and direction, and a hadronic jet with well-measured direction but poorly-measured
energy.  We will find that the three unknowns can be solved for using energy conservation and
momentum conservation in the plane of the event, allowing one to reliably reconstruct the energy of the neutrino.
Moreover, by imposing the requirement that the direction of the charged lepton, the hadronic jet, and the
direction of the source all lie in a plane, one can significantly reduce the background arising from
atmospheric neutrinos.

It is interesting to note that a similar idea was recently considered in a different context in Ref.~\cite{Kelly:2019wow}.
In that work, the authors considered the use of similar kinematic techniques to identify events with a
topology which is {\it not} consistent with neutrinos arriving from the Fermilab beamline; such events might
result from non-standard interactions in which neutrinos from the beamline interacted with the Fermilab
near detector to produce dark sector particles which are not identified.

This paper is structured as follows.  In Section~II, we describe the kinematic analysis we will use for
neutrino energy reconstruction.  In Section~III, we will apply this energy reconstruction algorithm to
a dark matter analysis.  In Section~IV, we determine the sensitivity which DUNE could achieve in a
search for monoenergetic neutrinos arising from dark matter annihilation in the Sun.  We conclude with a
discussion of our results in Section~V.

\section{Energy Reconstruction}

We will consider the case of (anti-)neutrinos arriving at a detector from a known direction, with energy $E_\nu \gtrsim 1\gev$.  We will focus on the case in which the incoming neutrino has a charged-current (CC) deep-inelastic
scattering (DIS) interaction with a quark or anti-quark in the nucleus.  Thus, we are interested in interactions
of the form $\nu_\ell + q_d (\bar q_u) \rightarrow \ell^- + q_u (\bar q_d)$, $\bar \nu_\ell + q_u (\bar q_d)
\rightarrow \ell^+ + q_d (\bar q_u)$, where $q_u$ and $q_d$ are up-type and down-type quarks, respectively.
The outgoing particles from this CC-DIS interaction are a charged lepton and a jet arising from the fragmentation
and hadronization of the outgoing quark.  We will consider the case in which $\ell = e$ or $\mu$.

\begin{figure}[ht]
\centering
\includegraphics[angle=0,width=0.5\textwidth]{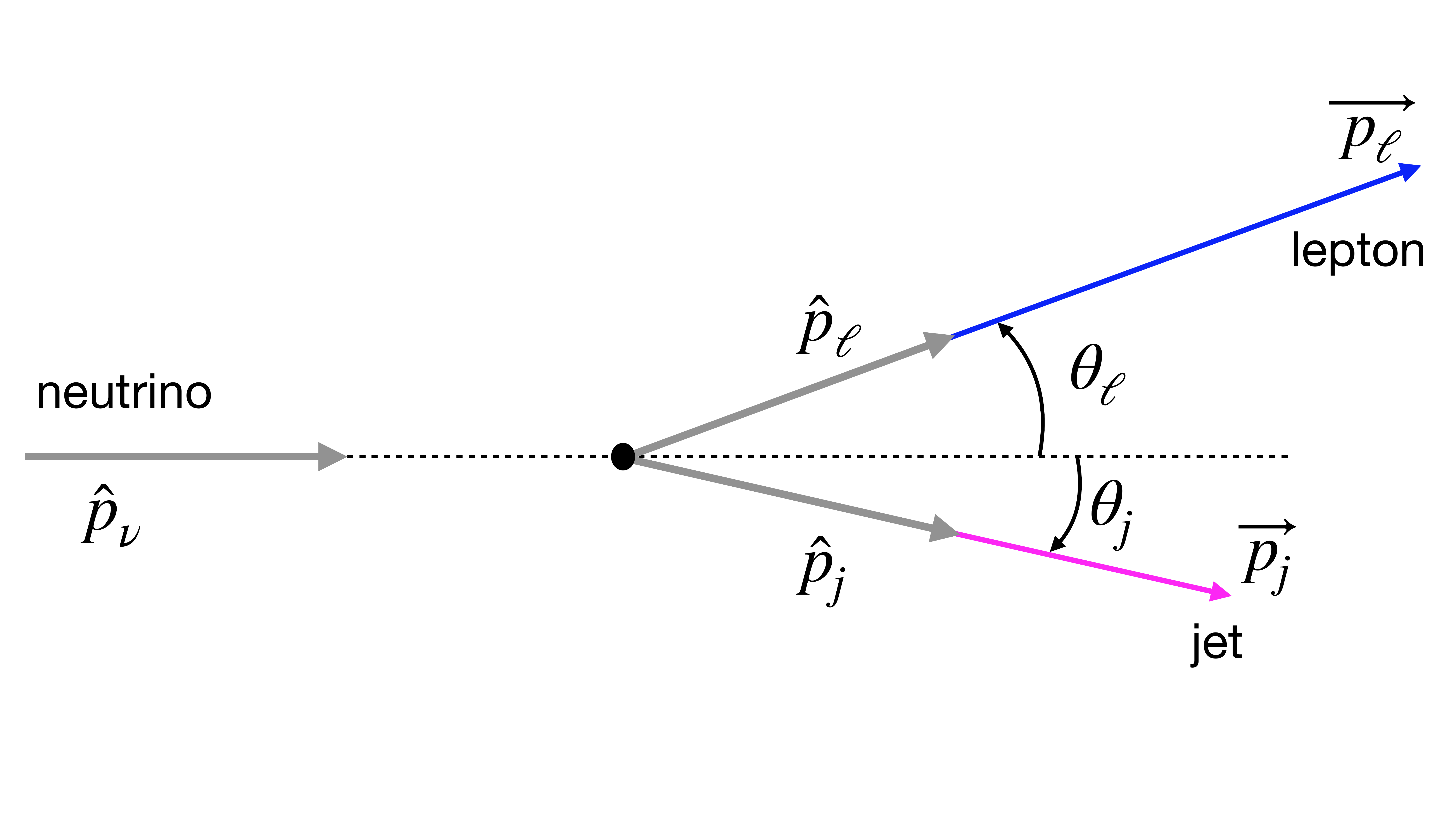}
\includegraphics[width=0.4\textwidth]{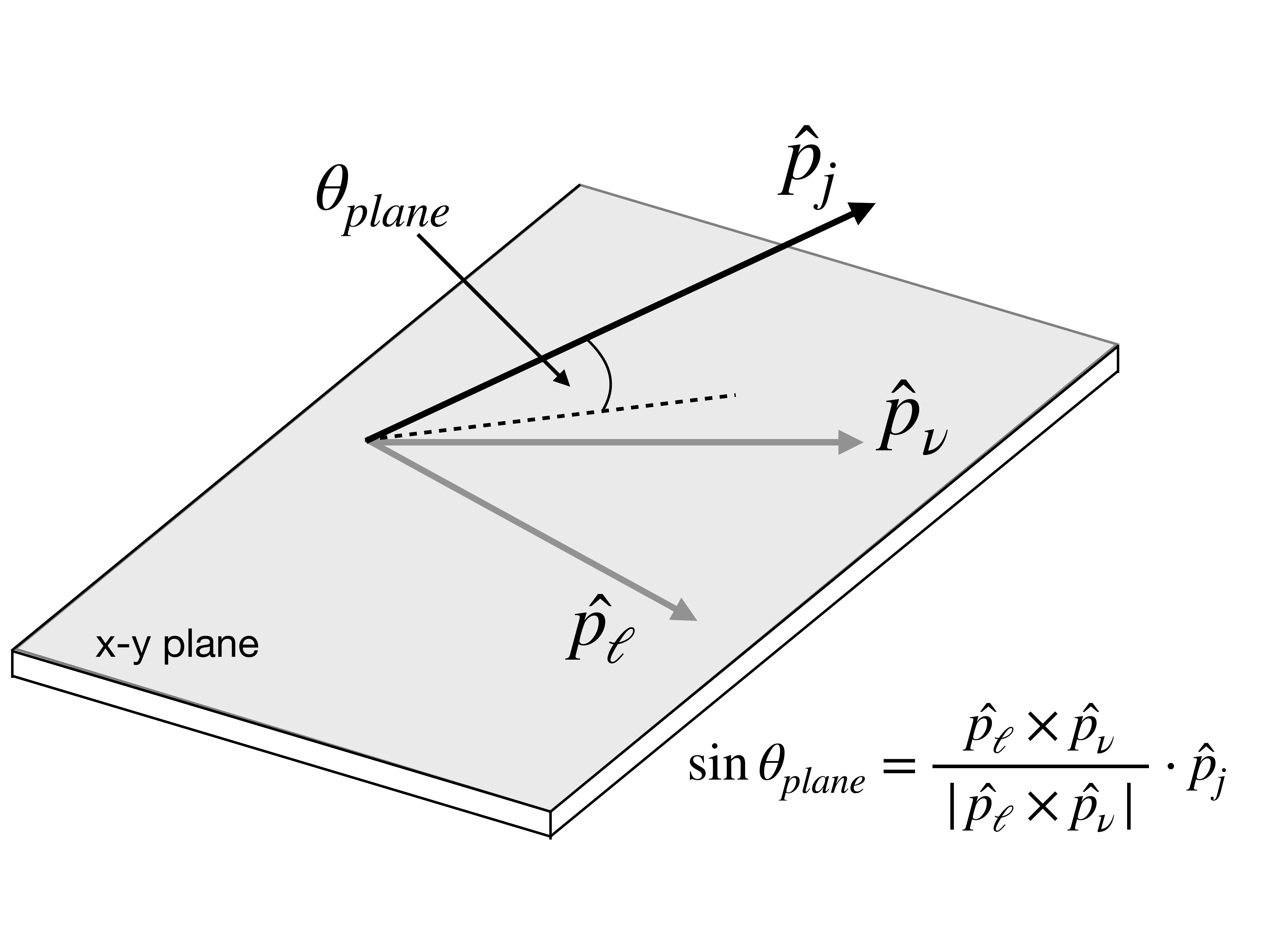}
\caption[DUNE]{Left: Definition of vectors used in this analysis based on the expected event topology following a deep-inelastic charged-current interaction with an argon nucleus in the detector, producing a charged lepton and a jet. Right: Definition of the ``$(x,y)$-plane" formed by the direction of the incoming neutrino and the lepton produced in the interaction. $\theta_{plane}$ represents the angle between the plane and the reconstructed jet.}
\label{fig:Topology}
\end{figure}

From momentum conservation, it is clear that the plane spanned by the momenta of the outgoing charged lepton, $\vec p_\ell$, and the outgoing jet, $\vec p_j$, must also include the momentum of the incoming neutrino.  We will refer to this as the $(x,y)$-plane, with the momentum of the incoming neutrino, $\vec p_{\nu}$, taken to be along the $x$-axis.  The outgoing charged lepton will be taken to have energy $E_\ell$, with momentum in the
$(x,y)$-plane at angle $\theta_\ell$ with respect to the $x$-axis (counterclockwise).  Similarly, the outgoing quark will hadronize to produce a jet of
energy $E_j$ in the $(x,y)$-plane at angle $\theta_j$ with respect to the $x$-axis (clockwise).  We illustrate this event
topology in Figure~\ref{fig:Topology} (left panel).
Due to uncertainties and incomplete track reconstruction, measured momentum vectors $\vec p_\nu$, $\vec p_\ell$, and $\vec p_j$ will not be coplanar in practice. For this reason, we will actually define the $(x,y)$-plane to be the plane spanned by $\vec p_\nu$ and $\vec p_\ell$ (since these are expected to be measured with good precision). To reject background events and poorly reconstructed jets, we then define $\theta_{plane}$ as the angle between the $(x,y)$-plane and $\vec p_j$, as illustrated in Figure~\ref{fig:Topology} (right panel).

We will assume that a detector such as DUNE can measure $\theta_\ell$ and $E_\ell$ with good accuracy, and can measure
the direction of the jet, $\theta_j$, from the tracks of the identified hadrons.  However, we will assume that $E_j$ is poorly measured
because some of the hadrons composing the jet will not deposit significant energy in the detector; thus, although the tracks
which are identified are enough to produce a good estimate for the direction of the jet, they do not yield a very accurate measurement
of the energy.

The unknown quantities in the scattering event are $E_\nu$, $E_j$, and $\sqrt{s}$, the center-of-mass energy of the scattering
hard process.  Note, $\sqrt{s}$ is not known a priori because the momentum fraction of the incoming parton is not known; equivalently,
we do not know the center-of-mass energy of the outgoing charged lepton and quark.  However, these unknown quantities are constrained by
energy conservation and momentum conservation in the $(x,y)$-plane.  Using these constraints, we can solve for $E_\nu$ in terms of
$E_\ell$, $\theta_\ell$ and $\theta_j$, yielding
\bea
E_\nu &=& \frac{1}{2}\frac{\sin \theta_j (1+\cos \theta_\ell) + \sin \theta_\ell (1 + \cos \theta_j)}{\sin \theta_j}  E_\ell .
\label{eq:NuEnergyRecon}
\eea
Note that, although $E_\ell$, $\theta_\ell$ and $\theta_j$ can be well-measured at DUNE, they can still only be measured
with some finite precision, leading to uncertainty in the reconstructed energy derived from Eq.~\ref{eq:NuEnergyRecon}.

Note also that if $\theta_j \rightarrow 0$, then $\theta_\ell \rightarrow 0$ as a result of momentum conservation in the plane
of the event.  Thus, Eq.~\ref{eq:NuEnergyRecon} is ill-defined if $\theta_{\ell,j} =0$.
Essentially, this amounts to saying that $E_j$ can only be determined by requiring the perpendicular component of the jet momentum
to cancel out the measured perpendicular component of the charged lepton momentum; if the jet and lepton are collinear then this method cannot be applied to determine $E_\nu$.
More generally, the uncertainty on reconstructed neutrino energy in Eq.~\ref{eq:NuEnergyRecon} becomes very large if the measured values of either $\theta_\ell$ or $\theta_j$ are very small.

We have thus far assumed that the momenta of the outgoing lepton and jet, along with a unit vector from the source to the detector, approximately lie in the same plane; this condition is required if the lepton and jet are indeed produced from a CC-DIS interaction initiated by a neutrino arriving from the putative source.  We can thus reject events which do not have this topology.  Since $\theta_{plane}$ measures deviations from this topology, we will reject events in which $\theta_{plane}$ is sufficiently far from zero.

\subsection{Numerical Simulation}

To estimate the precision with which we can reconstruct the neutrino energy, and the corresponding efficiency associated with the selection criteria, we simulate (monoenergetic) neutrino-argon scattering events
using the \verb+NuWro 17.01.1+ software package~\cite{Zmuda:2015twa}.  Our strategy for reconstructing the neutrino energy in this simulated sample will be as
follows:
\begin{itemize}
\item{For each event, obtain the charged lepton momentum ($\vec{p}_\ell$) and the jet direction ($\hat p_j$) from
the momentum vectors of the generated final state particles.}
\item{Smear $\vec{p}_\ell$ and $\hat p_j$ by the associated energy and angular resolutions of the detector. }
\item{Using these smeared vectors, determine $E_\ell$ and $\theta_{\ell}, \theta_j, \theta_{plane}$.}
\item{Apply cuts to $\theta_{\ell}, \theta_j, \theta_{plane}$, and obtain the fraction of simulated events which pass the cuts.}
\item{Reconstruct $E_\nu$ from $E_\ell$, $\theta_\ell$ and $\theta_j$ on an event-by-event basis using Eq.~\ref{eq:NuEnergyRecon}, and then use
the reconstructed $E_\nu$ distribution to determine
the mean, variance, and fraction of events contained within the variance. In principle, we project $\hat p_j$ onto the plane spanned by
$\vec p_\ell$ and $\vec p_\nu$, but since we will require $\theta_{plane} < 5^\circ$, the projection makes negligible difference.}
\end{itemize}
To implement this procedure, we will need the lepton
energy resolution, the lepton and jet angular resolutions, and an algorithm for determining $\hat p_j$.

The kinetic energy~(KE) thresholds used in our analysis are taken from the DUNE CDR~\cite{Acciarri:2015uup} and summarized in Table~\ref{tab:ThreshAndRes}. These thresholds are applied to the \verb+NuWro+-generated outgoing particles; only particles above these thresholds will be used to reconstruct the neutrino energy.
A full event reconstruction is beyond the scope of this work, for simplicity we consider particles as identified if they are above their respective detection threshold as defined in  Table~\ref{tab:ThreshAndRes}. Events are rejected if there is no lepton that falls above the  charged lepton kinetic energy threshold, or if none of the hadrons is above their respective energy threshold. In the rare case that multiple leptons fall above the energy threshold we also reject the events, though in practice one might use the leading lepton in these events. We assume that leptons and hadron can be reliably be identified, but the impact of misidentified leptons or hadrons should be studied separately and would require a full detector simulation and reconstruction. In summary we require events to have exactly one identified charged lepton and one or more identified hadrons. 
\begin{table}[h!]
\centering
\begin{tabular}{l c c c}
\hline
Particle Type ~~ & Detection Threshold (KE)\\
\hline \hline
$e^{\pm}/\mu^\pm/\gamma$  		& ~~30~MeV\\
$\pi^{\pm}$  	& ~100~MeV\\
$p/n/$other  	& ~~50~MeV\\
\hline
\end{tabular}
\caption{\label{tab:ThreshAndRes} Detection thresholds for various final state particles at DUNE, taken from \cite{Acciarri:2015uup}. These thresholds are applied to the generated events before our energy reconstruction algorithm is applied.}
\end{table}

Estimates for the lepton energy resolution $\epsilon_\ell$
can be found in~\cite{GrantYang}.  An electron will generally produce an electromagnetic shower
which will be contained within the detector ($\epsilon \sim 0.08$), while a muon will generally produce a track.  If the track
is contained within the detector, then the muon energy can be determined precisely ($\epsilon \sim 0.05$) from the length of the track.
If the track is only partially contained, however, then the energy can still be estimated, albeit with less precision, from the
deflection of the track due to small angle scattering of the muon as it passes through the
detector~\cite{Ankowski:2006ts,Abratenko:2017nki} ($\epsilon \sim 0.23$).
But these estimates were obtained for charged leptons with $E_\ell \leq 3~\gev$.  For  the neutrino energy range we are interested
in ($E_\nu \sim {\cal O}(1-10)\gev$) a significant fraction of the charged leptons will have $E_\ell > 3~\gev$, and higher
energy charged leptons typically have a better energy resolution.
Note also that for this neutrino energy range, one would expect most muons produced from CC-DIS interactions to exit the detector.
Similar energy resolution estimates can be found in~\cite{Acciarri:2015uup}.

A detailed estimate of the lepton energy resolution will ultimately require calibration of the actual detector, and the actual energy
resolution will likely depend on the orientation of the source relative to the detector.  For our purposes, a simple and reasonable
estimate is all that is required; we will assume that all $e^\pm$ produced by charged current interactions can be measured with an
energy resolution of $\epsilon_e = 5\%$, while $\mu^\pm$ can be measured with an energy resolution of $\epsilon_\mu =15\%$.

LArTPCs can track $e^\pm$ and $\mu^\pm$ with high precision and for our purposes we adopt a lepton angular resolution of $1^\circ$~\cite{Acciarri:2015uup}. Jet directional reconstruction is more complex and the directional angular resolution is expected to be significantly worse compared to single leptons. The development of a full
jet reconstruction algorithm is beyond the scope of this work. Hence, for our purposes we a adopt a straightforward reconstruction method that can be expected to give  reasonable and conservative results. It allows us to both reconstruct the jet direction and provides an uncertainty in the direction for events generated by \verb+NuWro 17.01.1+. Our method is based on the
momenta of the prompt particles in the hadronic jet which are above the detection threshold of the DUNE detector and hence could be observed.
Our procedure will be as follows:
the hadronic jet produced by the outgoing parton will consist of a set of various particles dominantly $p$, $n$, $\pi^\pm$, and $\pi^0$. While protons, neutrons, and charged pions can be detected individually by DUNE, neutral pions will decay and be detected through their di-photon signature.   For a given \verb+NuWro+ event with $N$ particles above the DUNE detection threshold~\cite{Acciarri:2015uup}, each with momentum $\vec{p}_i$,
we define the direction of the jet with the unit vector $\hat p_j$ given by

\bea
\vec p_{j} &=& \sum_{i=1}^N \vec{p}_i
\nonumber\\
\hat p_{j}  &=& \frac{\vec  p_{j}}{\left| \vec  p_{j}  \right|}.
\eea

We do not reconstruct neutral pions separately, but include the photon momenta from neutral pion decays in the computation of the total jet momentum, if the respective photon is above its detection threshold.
At DUNE beam energies we can expect that $\pi^0$'s can in general be identified through their di-photon signatures, which is significantly more difficult for other experiments~\cite{Abe:2017urf}.

Two sources of uncertainty enter in our estimate of the jet direction.  First,  some of the particles produced through hadronization and fragmentation
from the initial outgoing parton are missed, if their energy is below the energy threshold for particle identification.  Secondly, the energy and direction of
the observed particles in the hadronic jet will be distorted due to the detector energy and angular resolution.  The former source of uncertainty is accounted
for by reconstructing the jet direction using only particles whose energy lies above the detection thresholds estimated in~\cite{Acciarri:2015uup}.  We will
account for the latter source of uncertainty by smearing $\hat p_j$ by $5^\circ$.  The angular resolutions for the particles which dominate the hadronic cascade
are expected to lie in the $1^\circ - 5^\circ$ range~\cite{Acciarri:2015uup}, so our estimate is conservative.  We have also considered more optimistic choices
of $1^\circ$ and $3^\circ$ for the jet angular resolution, and find marginal change in the overall results.

Finally, having obtained the smeared values of $\vec{p}_\ell$ and $\hat p_j$, we find
\bea
E_\ell &=& \sqrt{\vec{p}_\ell^{~2} + m_\ell^2} \sim \left| \vec{p}_\ell \right| ,
\nonumber\\
\cos \theta_\ell &=& \hat p_{\nu} \cdot \hat p_\ell ,
\nonumber\\
\cos \theta_j &=& \hat p_{\nu} \cdot \hat p_j ,
\nonumber\\
\sin \theta_{plane} &=& \left| \frac{ \hat p_\ell \times \hat p_{\nu} }{ \left| \hat p_\ell \times \hat p_\nu \right|} \cdot \hat p_j \right|.
\eea

In addition, we define a variable $E_j^{dep.}$, which is an estimate of the amount of jet energy deposited in the
detector:
\bea
E_j^{dep.} &\equiv& (m_{\rm remnant~nucleus} - m_{\rm initial~nucleus}) + \sum_{i=1}^{N} \sqrt{|\overrightarrow{p}_i^2| + m_i^2},
\eea
where $N$ counts the number of hadronic tracks above our detection thresholds.
This quantity will typically underestimate the actual energy of the jet, since some of the hadrons composing the jet will fall
below the energy threshold. We therefore do not use $E_j^{dep.}$ in reconstructing the neutrino energy, and instead utilize the
algorithm we have previously described.  However, $E_j^{dep.}$ is a useful variable on which to apply selection cuts, in order to reject
events in which insufficient jet energy is detected.
In particular note that, in defining $E_j^{dep.}$, we subtract the difference in mass between the initial argon nucleus and the final remnant nucleus.  This essentially accounts for the mass of any nucleons which are ejected by final state interactions.  As a result,
for events in which very little jet energy is contained in hadrons above threshold, we will find $E_j^{dep.} \leq 0$.  For these events,
since so little of the jet energy can be detected, it is likely that jet direction will also be reconstructed inaccurately.  We will
therefore require $E_j^{dep.} >0$.

\begin{figure}[ht]
\centering
\includegraphics[width=.8\textwidth]{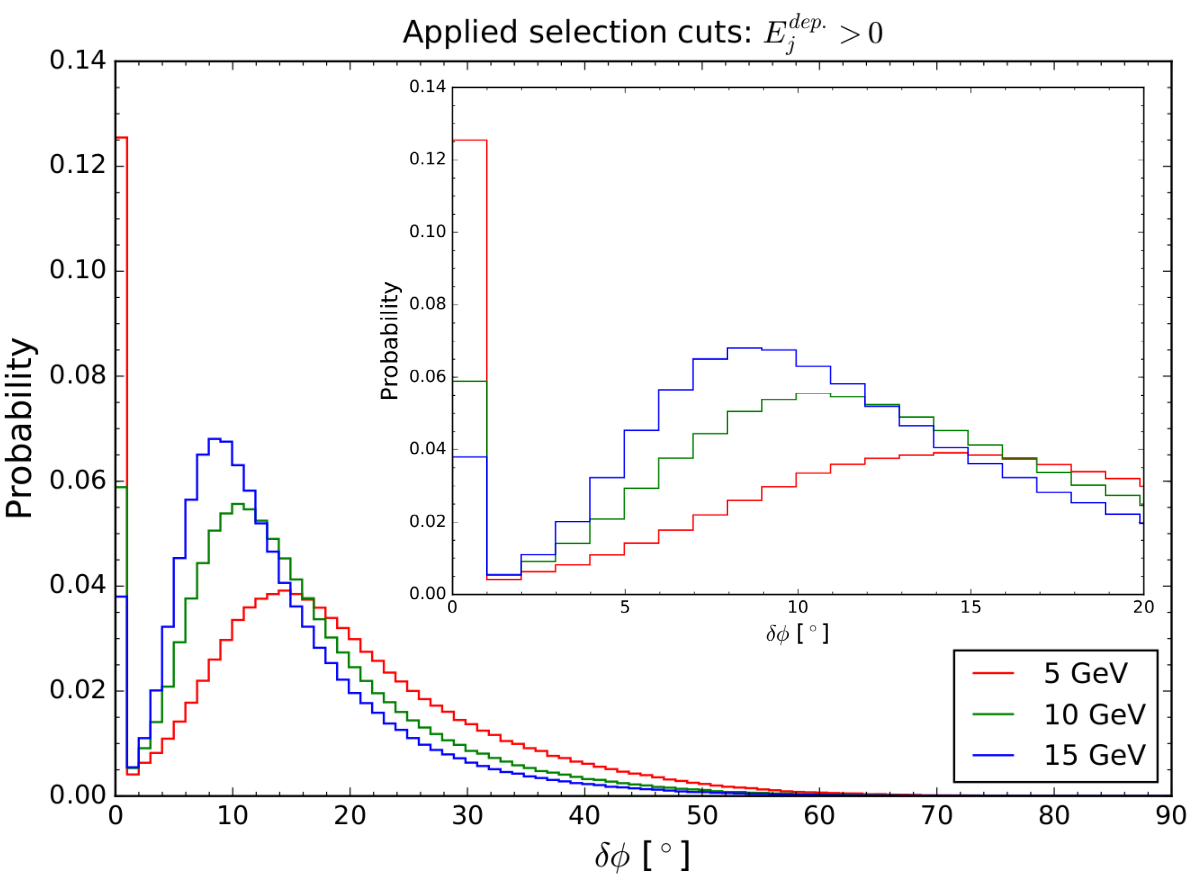}
\caption{Probability distribution of $\delta \phi$ for $5\gev$, $10\gev$, and $15\gev$ incoming muon neutrinos.  We do not include any smearing to account for the energy or
angular resolution of the detector. The peak at zero is caused by jets with only one particle above the reconstruction energy threshold. }
\label{fig:DeltaPhi}
\end{figure}

We also define the variable
\bea
\delta \phi &\equiv& \frac{\sum_i^N |\vec{p}_i | \sin \phi_i}{|\sum_i^N \vec{p}_i |} ,
\eea
where $\phi_i$ is the positive angle between $\vec{p}_i$ and $\hat p_j$.  $\delta \phi$ characterizes
the angular size of the visible jet resulting from hadronization and fragmentation of the outgoing parton. We compute $\delta \phi$ by taking the momentum-weighted sum of the angular distance of the jet constituents above their respective detection thresholds.
For illustration we plot the
distribution of $\delta \phi$ arising from CC-interactions of $5\gev$, $10\gev$, and $15\gev$ muon neutrinos
(for events with $E_j^{dep.} >0$) in Figure~\ref{fig:DeltaPhi}.
We note that the large peak in the $\delta \phi$ probability distribution near $\delta \phi =0$ arises from events where only one
hadron in the jet is visible and above the detection threshold. In Figure~\ref{fig:DeltaPhi}, we
have not included any smearing of the visible hadron momenta to account for the energy or angular resolution of the detector.
However, one expects the effect of the angular resolution to be negligible, since the intrinsic size $\delta \phi$ is significantly
larger than the estimated angular resolution of the individual visible hadrons composing the jet. For this reason, we would expect
that the neutrino energy resolution of our reconstruction algorithm should be largely independent of our choice of angular resolution
for the individual hadron tracks, since this uncertainty will in any case be much smaller than the intrinsic size of the jet. Indeed, as noted earlier, utilizing more optimistic hadron track angular resolutions (of $1^\circ$ and $3^\circ$) has a marginal effect on our final results.

\section{Energy Reconstruction for a Dark Matter Analysis}

We define a set of selection criteria based on basic event reconstruction principles and perform a selection cut optimization for a dark matter analysis to determine the efficiency and acceptance of our energy reconstruction procedure. For a search for neutrinos
arising from dark matter annihilation in the Sun, the cuts should be designed to maximize the energy resolution and acceptance of events in which the neutrino arrives from the direction
of the Sun, while maximizing the rejection of background atmospheric neutrino events in which the neutrino does not arrive from the direction of the Sun.  The guiding
considerations we will find are:
\begin{itemize}
\item{An upper bound on $\theta_\ell$ and/or $\theta_j$: -- CC-DIS events involving a neutrino from the Sun will
typically produce a lepton and a jet pointing away from the Sun (small $\theta_\ell$, $\theta_j$).}
\item{A lower bound on $\theta_\ell$ and/or $\theta_j$: --  if $\theta_{\ell}$ and $\theta_{j}$ are very small, then the jet is nearly collinear with the lepton, and hence the uncertainty in the reconstructed neutrino energy as determined from Eq.~\eqref{eq:NuEnergyRecon} becomes very large.}
\item{An upper bound on $\theta_{plane}$: -- if $\theta_{plane}$ is not small, then the plane spanned by the lepton and jet momenta does not contain the Sun, implying that topology of the event
is inconsistent with a neutrino originating from the Sun.}
\end{itemize}

We optimize our choice of selection cuts by maximizing the signal-to-background ratio, while at the same time finding a set of cuts that have a broad signal acceptance over a large range of dark matter masses.  To determine the choice of
cuts, we have performed the neutrino energy reconstruction procedure on simulations. We generated 48 samples, each consisting of $10^6$ monoenergetic CC neutrino events. For each flavor ($\nu_e$, $\bar \nu_e$, $\nu_\mu$, and $\bar \nu_\mu$)  samples were generated in $1\gev$ increments from 4 to $15\gev$.  These samples represent signal events in which dark matter in the Sun annihilates to produce
monoenergetic neutrinos via the process $\chi \chi \rightarrow \nu_\ell \bar \nu_\ell$.  In addition, four background samples of $10^6$ $\nu_e$, $\bar \nu_e$, $\nu_\mu$ or $\bar \nu_\mu$ CC-events
are generated.  In each of these background samples, the energies of the $\nu_e$, $\bar \nu_e$, $\nu_\mu$ or $\bar \nu_\mu$ are given a distribution between the energies
$E_{min}=4\gev$ and $E_{max}=20\gev$ consistent with the atmospheric
neutrino background
at the location of Homestake\footnote{Note that the flux normalizations are indeed as given Eq.~\ref{eq:HomestakeFlux} and not a typo.}:
\bea
{d\Phi_{atm.}^{\nu_e} \over dE} &\sim& 4\pi (398~\m^{-2} \s^{-1} \gev^{-1} ) \times \left(0.53 + \frac{E_\nu}{\gev}\right)^{-3.58} ,
\nonumber\\
{d\Phi_{atm.}^{\bar \nu_e} \over dE} &\sim& 4\pi  (398~\m^{-2} \s^{-1} \gev^{-1} ) \times \left(0.60 + \frac{E_\nu}{\gev}\right)^{-3.74} ,
\nonumber\\
{d\Phi_{atm.}^{\nu_\mu} \over dE} &\sim& 4\pi  (398~\m^{-2} \s^{-1} \gev^{-1} ) \times \left(0.34 + \frac{E_\nu}{\gev}\right)^{-3.10} ,
\nonumber\\
{d\Phi_{atm.}^{\bar \nu_\mu} \over dE} &\sim& 4\pi  (278~\m^{-2} \s^{-1} \gev^{-1} ) \times \left(0.21 + \frac{E_\nu}{\gev}\right)^{-2.98}.
\label{eq:HomestakeFlux}
\eea
These functional forms are obtained by a fit to the angle-averaged solar minimum flux values tabulated in~\cite{AtmNuBgd}
(referenced in the published article~\cite{Honda:2015fha}).

We will choose a single set of cuts for $\theta_{\ell}, \theta_{j}, \theta_{plane}$, and determine the
mean and variance of the reconstructed neutrino energy for the simulated sample of signal events which pass the cuts,
for each choice of true (anti-)neutrino energy in the $4-15$ GeV range.
In all cases we will find that
the difference between the mean of the reconstructed neutrino energy distribution and the true neutrino energy is much smaller than the variance.  When
searching for neutrinos with a particular true energy, we will thus look for events with a reconstructed neutrino energy in a window
centered at the desired true neutrino energy, with a width given by $\pm 1\sigma$ variance.

The selection criteria are chosen to minimize background, while maximizing signal acceptance.  For simplicity, we have chosen one set of cuts for
a search for neutrinos of any energy in the $4-15$ GeV range.
The total number of expected background events (before the application of cuts) is given by
\bea
N_{bgd}^{\nu_\ell, \bar \nu_\ell} &=& \int_{E_{min}}^{E_{max}} dE ~{d \Phi_{atm.}^{\nu_\ell, \bar \nu_\ell} \over dE} \times  \left( T A^{\rm eff.}_{\nu_\ell, \bar \nu_\ell} \right)
\nonumber\\
&=& \left( \frac{\rm exposure}{\kT ~\yr} \right)
\left(1.9 \times 10^{40}  \right)
 \int_{E_{min}}^{E_{max}} dE ~\left( {d \Phi_{atm.}^{\nu_\ell, \bar \nu_\ell} \over dE}(\cm^2~\s)  \right) \times  \frac{\sigma_{\nu, \bar \nu N}}{\cm^2} ,
\eea
where the effective area, $A^{\rm eff.}_{\nu_\ell, \bar \nu_\ell}$, is the product of the number of target nucleons and the
(anti-)neutrino/nucleon CC-DIS scattering
cross sections. We use the CC-DIS cross section as computed by \verb+NuWro+ (given in appendix~\ref{app:cross}), which is in good agreement to the scaling approximation~\cite{Brodsky:1997yr}
\bea
\sigma_{\nu N} &\sim& (6.9 \times 10^{-39} \cm^2) \left({E \over \gev} \right) ,
\nonumber\\
\sigma_{\bar \nu N} &\sim& (3.2 \times 10^{-39} \cm^2) \left({E \over \gev} \right) .
\eea
We will assume that the effective exposure of DUNE for this analysis is 380~kT~yr.
Given this exposure and the corresponding total number of background events, we have adopted a set of cuts which would yield $\lesssim {\cal O}(1)$ expected background event surviving
the angular cuts and lying within any one reconstructed energy window.

\begin{figure}[ht]
\centering
\includegraphics[width=0.8\textwidth]{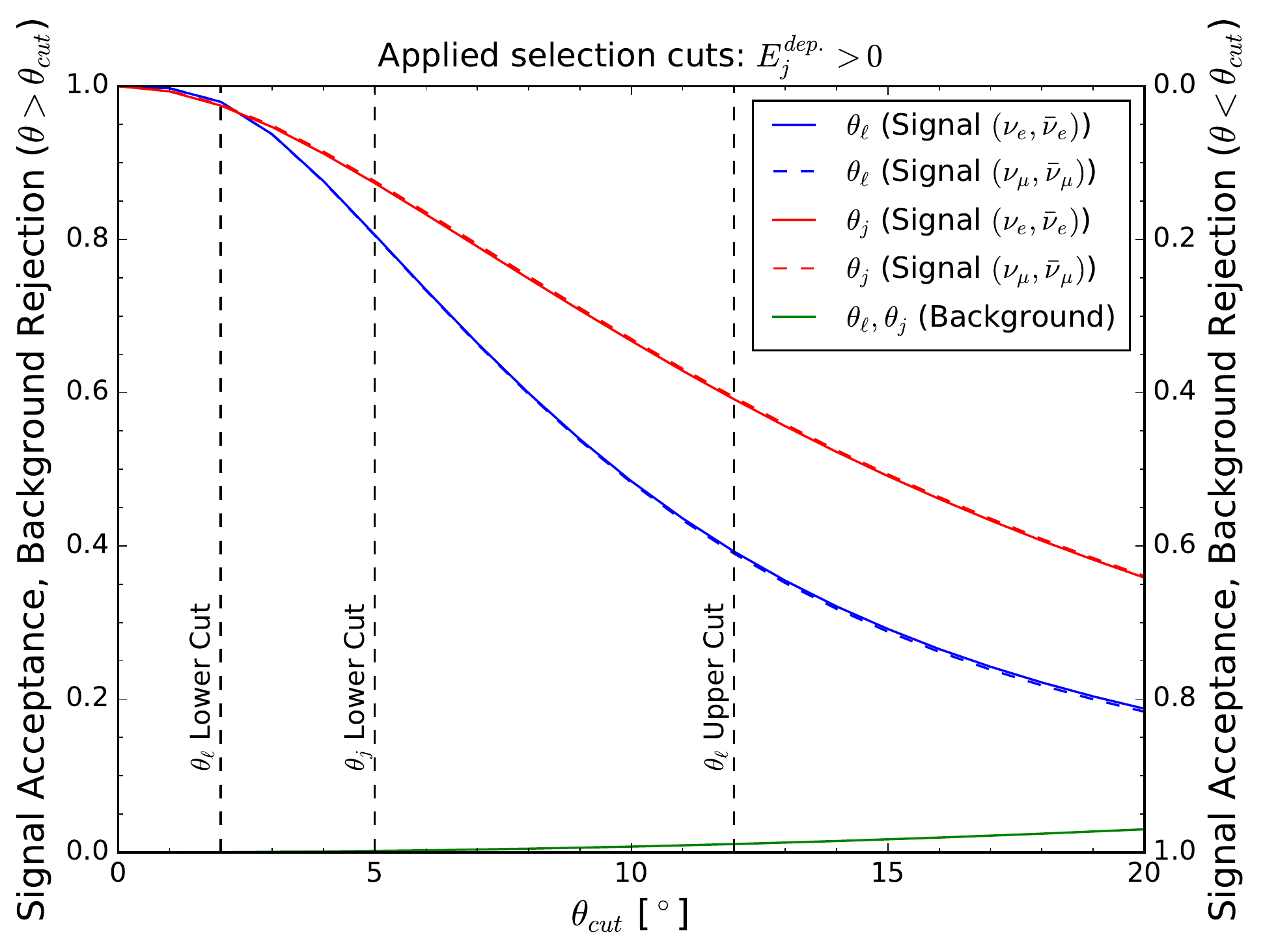}
\caption{Plots of the fraction of signal events which pass cuts $\theta_\ell > \theta_{\ell, cut}$ (left axis, blue), $\theta_\ell < \theta_{\ell, cut}$ (right axis, blue), $\theta_j > \theta_{j, cut}$ (left axis, red) and $\theta_\ell < \theta_{j, cut}$  (right axis, red), in the case of
$10\gev$ (anti-)neutrinos arriving from the direction of the Sun with equal fluxes of the
$\nu_e, \bar \nu_e, \nu_\mu, \bar \nu_\mu$. Solid lines represents the $(\nu_e, \bar \nu_e)$ channels, and dashed lines the $(\nu_\mu, \bar \nu_\mu)$ channels.
The fraction of atmospheric background events rejected by the same cuts are plotted in green (the background rejection
curves for leptons and jets are identical, and independent of energy or \mbox{(anti-)}neutrino flavor).
We have not accounted for the angular resolution with any smearing of the lepton or jet direction.}
\label{fig:HistFractionPassingAngleCuts}
\end{figure}

To illustrate how the signal acceptance changes with the choice of angular cuts, we plot in Figure~\ref{fig:HistFractionPassingAngleCuts} the fraction of accepted CC events ($E_j^{dep.} >0$) with a (anti-)neutrino
arriving from the Sun with true energy $E_\nu =10\gev$ for which $\theta_\ell > \theta_{\ell,cut}$ (left axis, blue) or $\theta_\ell < \theta_{\ell,cut}$ (right axis, blue) and $\theta_j > \theta_{j,cut}$ (left axis, red) or $\theta_j < \theta_{j,cut}$ (right axis, red).
We assume equal fluxes of $\nu_e, \bar \nu_e, \nu_\mu, \bar \nu_\mu$.
We also plot the fraction of atmospheric neutrino background events rejected by cuts $\theta_\ell > \theta_{\ell,cut}$ (left axis, green)
and $\theta_\ell < \theta_{\ell,cut}$ (right axis, green).  Since we approximate the average atmospheric neutrino background to
be isotropic, the fraction of background events rejected is independent of energy and (anti-)neutrino flavor,
and cuts on the jet direction produce an identical
background rejection curve.
We have not accounted for the angular resolution with any smearing of the lepton or jet direction.
The black dashed lines in Figure~\ref{fig:HistFractionPassingAngleCuts} represent the selection cuts ($2^\circ < \theta_\ell < 12^\circ$, $\theta_j > 5^\circ$) for this analysis as determined by a numerical survey.

\begin{figure}[ht]
\centering
\includegraphics[width=0.8\textwidth]{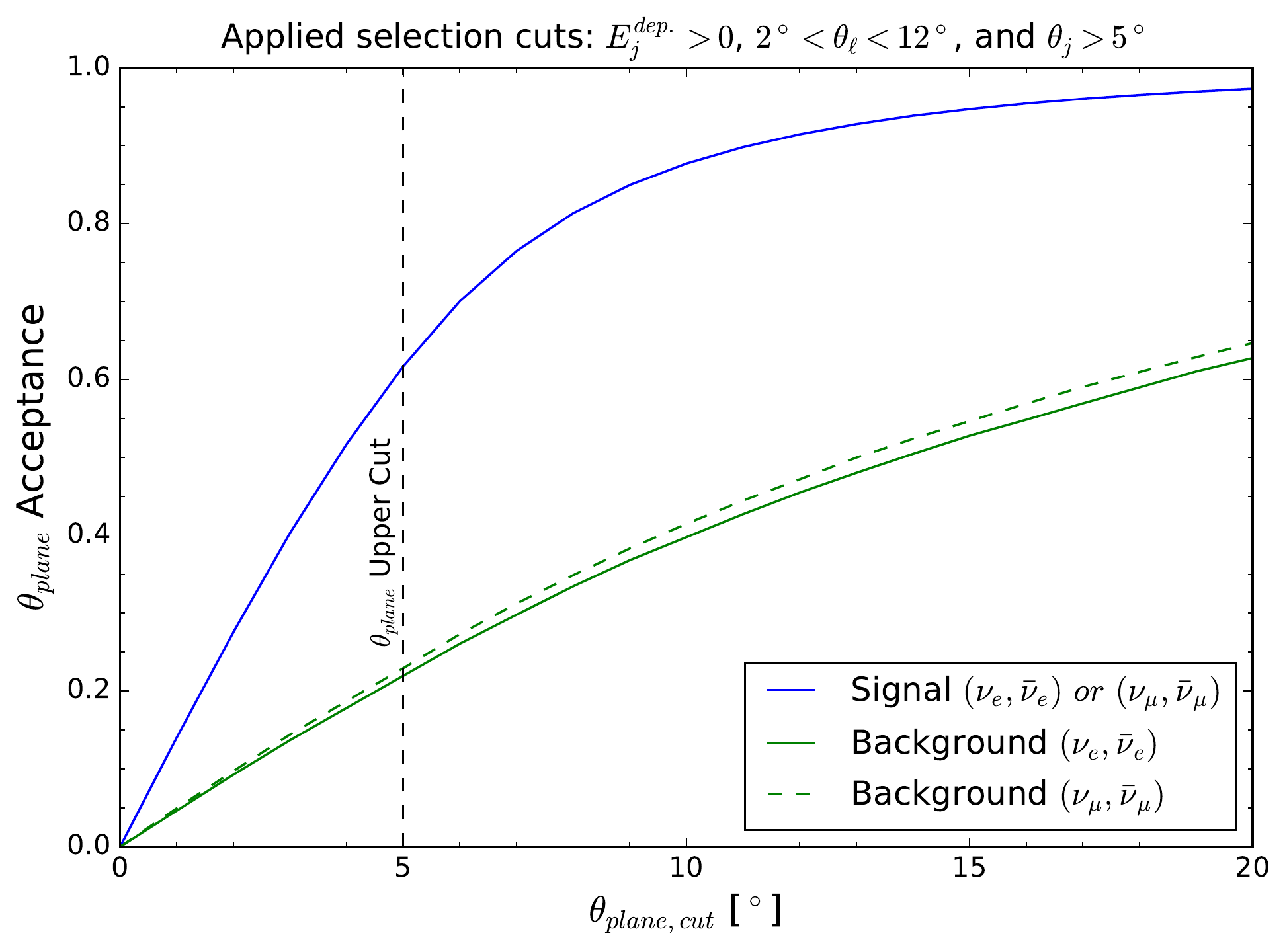}
\caption{The fraction of events that pass the selection criteria $\theta_{plane} < \theta_{plane,cut}$,
for $10\gev$ (anti-)neutrinos arriving from the direction of the Sun (blue), and for the atmospheric neutrino background of $(\nu_e, \bar \nu_e)$ (solid green) and $(\nu_\mu, \bar \nu_\mu)$ (dashed green).
We show the combined signal events from all channels assuming that dark matter annihilation produces equal fluxes of $\nu_e, \bar \nu_e, \nu_\mu, \bar \nu_\mu$. For the background we separate the acceptance for $(\nu_e, \bar \nu_e)$ and $(\nu_\mu, \bar \nu_\mu)$ channels, it differs due ot the different neutrino to anti-neutrino ratios on the background. The vertical black dashed line represents the $\theta_{plane}$ cut ($\theta_{plane} < 5^\circ$) that has been chosen for this analysis.}
\label{fig:ThetaPlane}
\end{figure}

Similarly, in Figure~\ref{fig:ThetaPlane}, we plot the fraction of events with $E_j^{dep.} >0$,
$2^\circ < \theta_\ell < 12^\circ$, $\theta_j > 5^\circ$ for which $\theta_{plane} < \theta_{plane,cut}$.
We plot signal events with true (anti-)neutrino energy of $10\gev$ in blue, and atmospheric neutrino background events in solid green for $(\nu_e, \bar \nu_e)$ and in dashed green for $(\nu_\mu, \bar \nu_\mu)$.
Again, we assume that dark matter annihilation produces equal fluxes of $\nu_e, \bar \nu_e, \nu_\mu, \bar \nu_\mu$. This figure illustrates how the signal and background acceptance changes as one changes the cut on $\theta_{plane}$.
The $(\nu_e, \bar \nu_e)$ and $(\nu_\mu, \bar \nu_\mu)$ channels result in apparently the same acceptance curve,
because the only difference that affects the angular distribution is the mass of the lepton, which is small compared to the energy scale here. In addition, the angular distribution is different for neutrino/nucleon scattering versus anti-neutrino/nucleon scattering.
The discrepancy between the background acceptance of $(\nu_e, \bar \nu_e)$ and that of $(\nu_\mu, \bar \nu_\mu)$ occurs because the relative number of neutrinos versus anti-neutrinos is different, calculated according to Eq.~\ref{eq:HomestakeFlux}. However, if no angle cuts on $\theta_\ell$ and $\theta_j$ have been applied, the acceptance of the background would be one identical curve.
The black dashed line in Figure~\ref{fig:ThetaPlane} represents the optimized cut ($\theta_{plane} < 5^\circ$) for this analysis.

\begin{figure}[ht]
\centering
\includegraphics[width=0.49\textwidth]{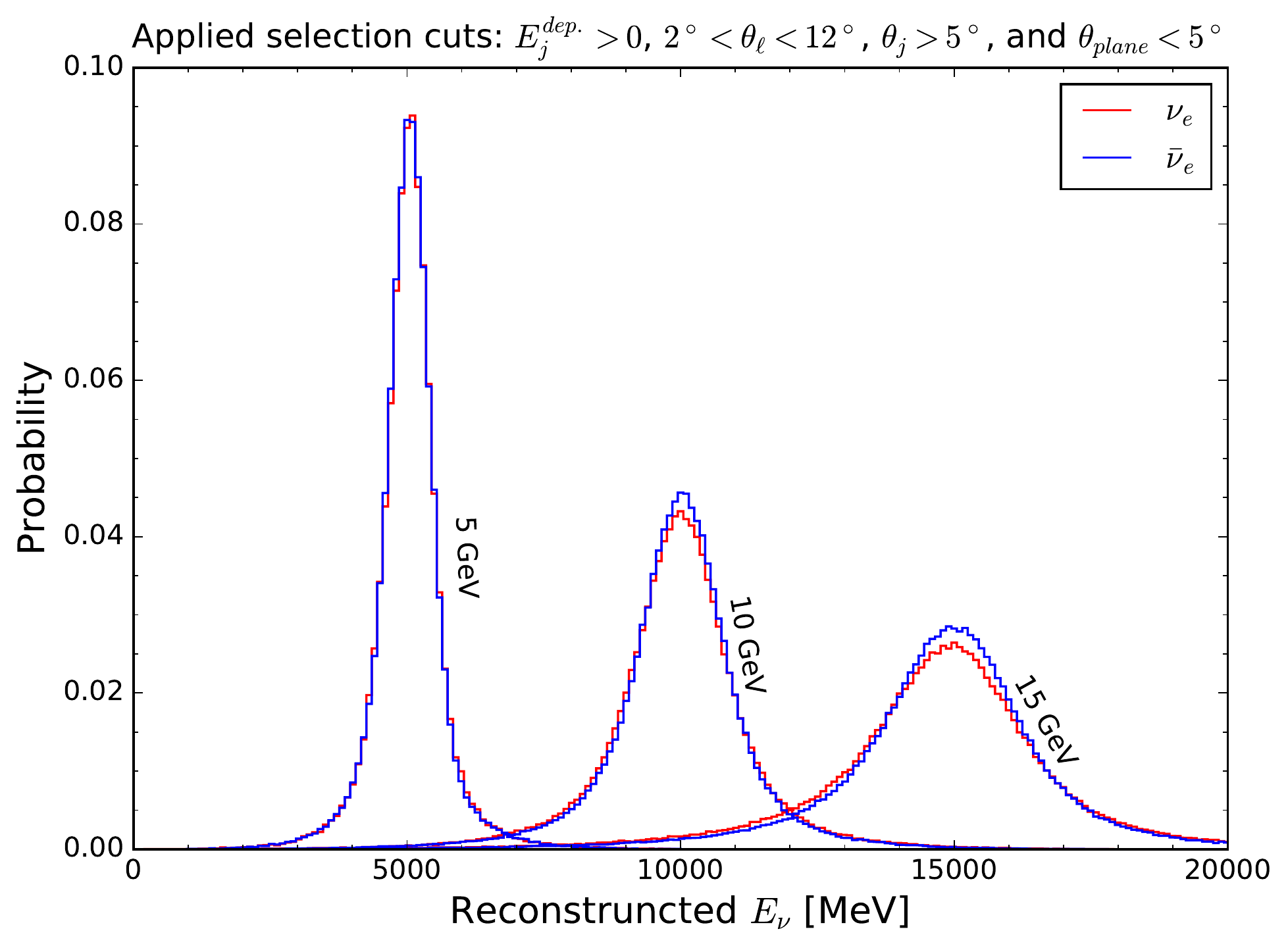}
\includegraphics[width=0.49\textwidth]{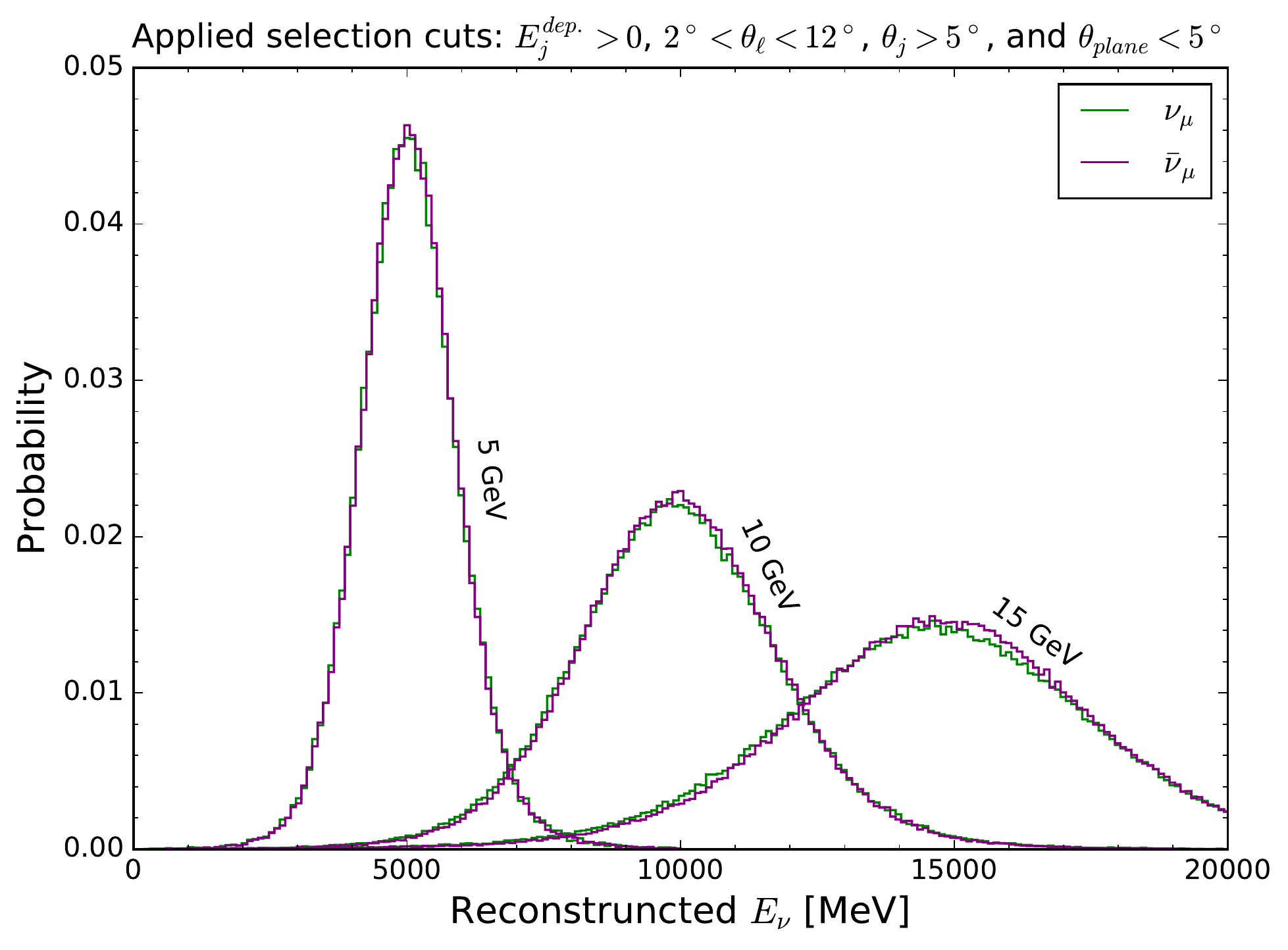}
\caption{The reconstructed neutrino energy for $\bar \nu_e, \nu_e$ (left panel) and
$\bar \nu_\mu, \nu_\mu$ (right panel) for true energies of 5, 10, and 15~GeV, as indicated.
Selection criteria in Table~\ref{tab:AngleCuts} are applied.  Since the electron is generally contained in the LArTPC,
the resulting reconstructed energy resolution is higher, while the muon is typically exiting and has a lower
energy resolution.}
\label{fig:ReconstructedEnergy}
\end{figure}

\begin{figure}[ht]
\centering
\includegraphics[width=0.8\textwidth]{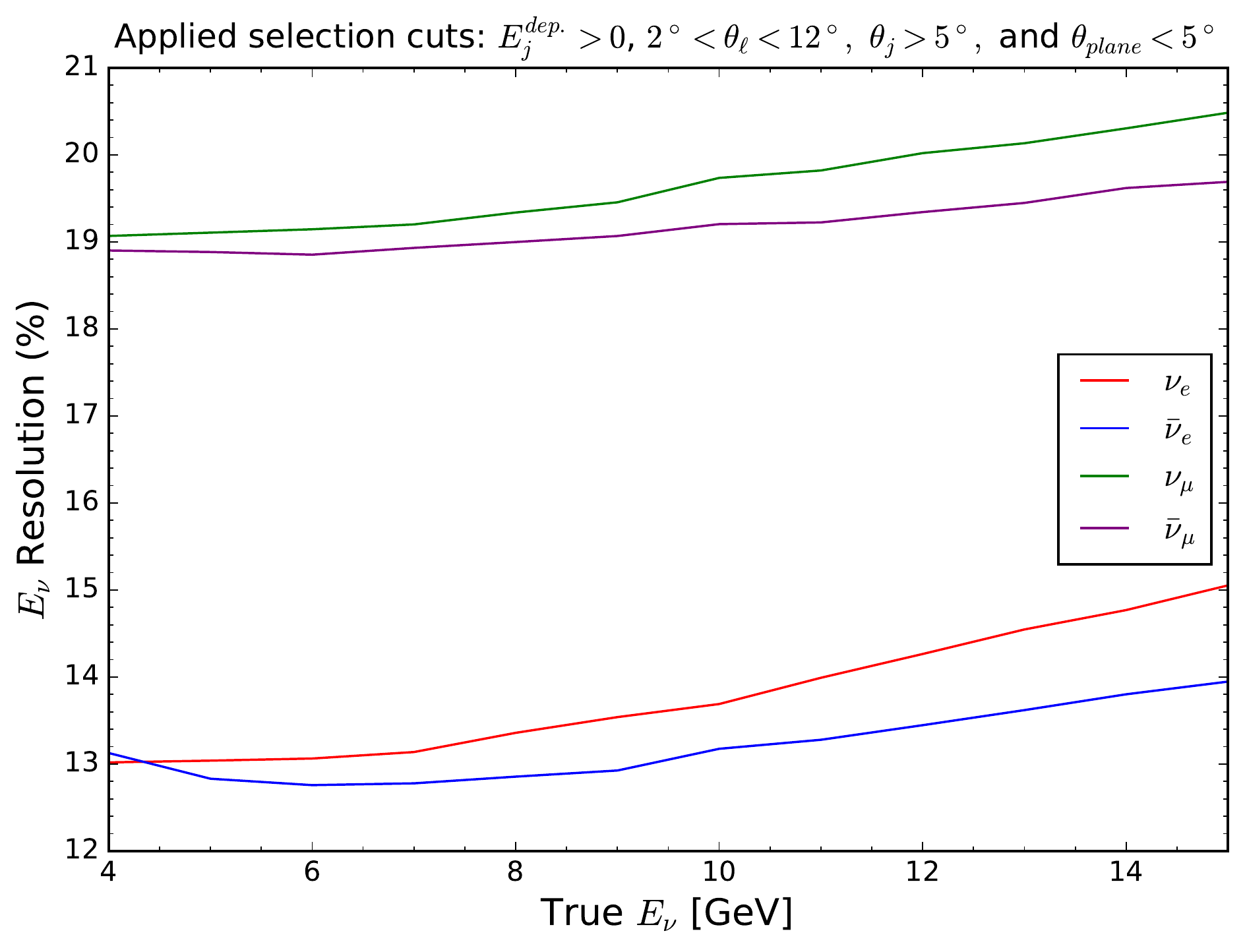}
\caption{Plots of the energy resolution ($1\sigma$ variance) as a function of the true neutrino energy, for
the $\nu_e$ (red), $\bar\nu_e$ (blue), $\nu_\mu$ (green),  and $\bar\nu_\mu$ (purple)
channels.}
\label{fig:EnergyResolution}
\end{figure}

We summarize our choice of selection cuts in Table~\ref{tab:AngleCuts}, along with the resulting signal acceptance (for $E_\nu = 10\gev$) and background rejection efficiencies.  After applying these cuts, the probability distribution of the reconstructed neutrino energy is plotted in Figure~\ref{fig:ReconstructedEnergy}
for the cases where the true neutrino energy is 5~GeV, 10~GeV, and 15~GeV.  The left panel of this figure corresponds to the ($\nu_e,\bar\nu_e$) channels while the right panel corresponds to the ($\nu_\mu,\bar\nu_\mu$) channels.  In Figure~\ref{fig:EnergyResolution} we plot the variance of these probability distributions as a function of the true neutrino energy for the $\nu_e$ (red), $\bar\nu_e$ (blue), $\nu_\mu$ (green), and $\bar\nu_\mu$ (purple) channels.
The energy resolution is roughly $\sim 20\%$ for the muon channels, in which the charged lepton
typically exits the detector.  However, it is roughly $\sim 15\%$ for the electron channels, in which the charged
lepton is typically contained; in this case, the uncertainty in the neutrino energy is much larger than the uncertainty
in the charged lepton energy, as it is dominated by the uncertainty in the jet direction.  It is interesting to
note that both of these results are comparable to those determined by~\cite{GrantYang}, using the deposited energy.

\begin{table}[ht!]
\centering
\begin{tabular}{|c|c|c|}
  \hline
  Event selection criteria & Signal acceptance & Background rejection \\
  \hline
  One $\ell$ and $j$ identified & & \\
  $E_j^{dep.} >0$ & 27.06 \% $(\nu_e,\bar\nu_e)$ & 99.88 \% $(\nu_e,\bar\nu_e)$ \\
  $2^\circ < \theta_\ell < 12^\circ$ & & \\
  $\theta_j > 5^\circ$ & 27.13 \% $(\nu_\mu,\bar\nu_\mu)$ & 99.87 \% $(\nu_\mu,\bar\nu_\mu)$ \\
  $\theta_{plane} < 5^\circ$ & & \\
  \hline
\end{tabular}
\caption{The event selection criteria used in our study, along with the resulting signal acceptance (for $E_\nu = 10\gev$) and background rejection efficiencies, are summarized in this table. The lower bound of $\theta_\ell$ is chosen based on Figure~\ref{fig:HistFractionPassingAngleCuts} such that the reconstructed energy is well-defined, as discussed below Eq.~\ref{eq:NuEnergyRecon}. The other bounds are chosen to reduce the background and increase the signal-background ratio.}
\label{tab:AngleCuts}
\end{table}

The event selection criteria used in this analysis, along with the resulting signal acceptance (for $E_\nu = 10\text{ GeV}$)  and background rejection efficiencies.

\begin{figure}[ht]
\centering
\includegraphics[width=0.8\textwidth]{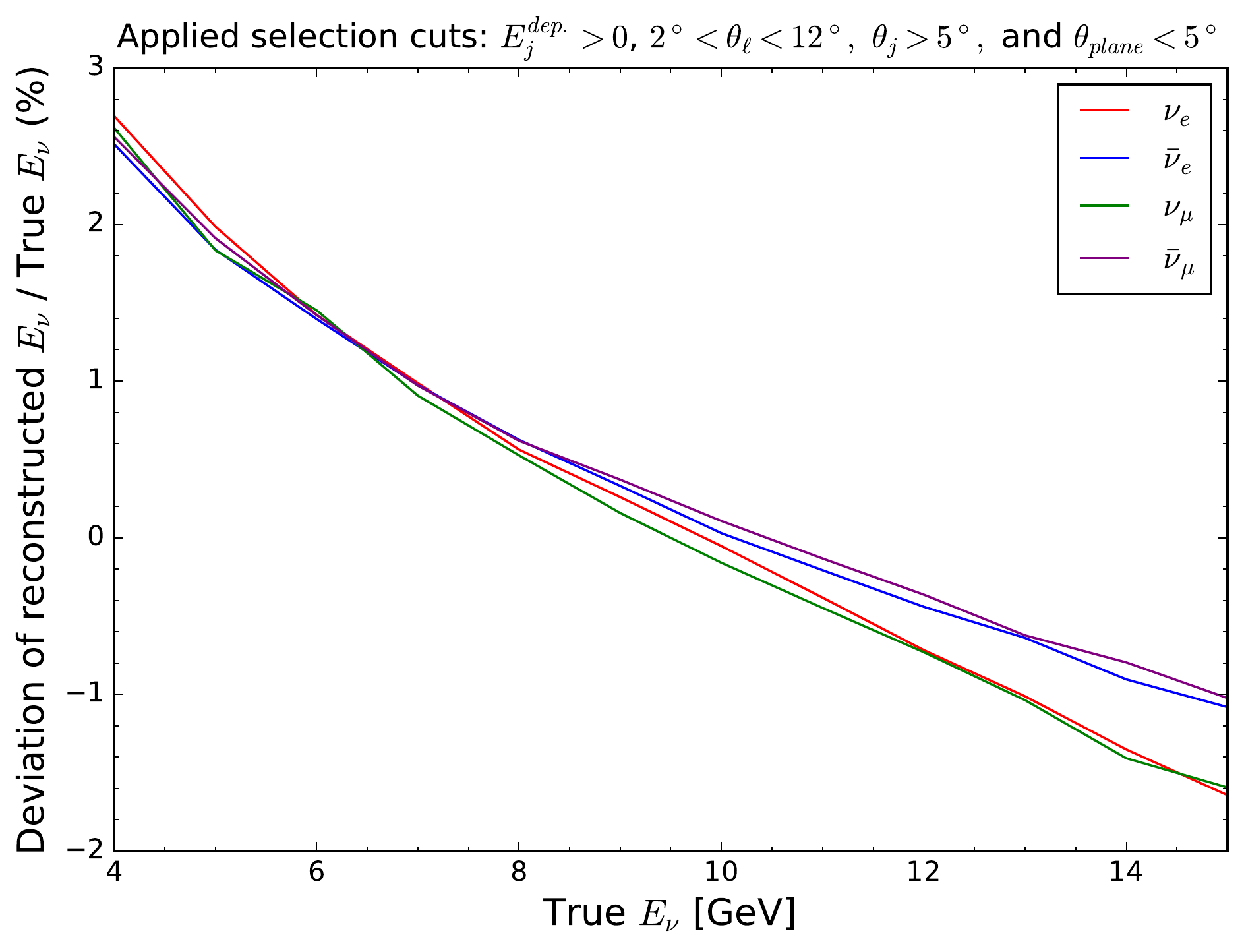}
\caption{Plots of the fractional deviation of the reconstructed neutrino energy from the true neutrino energy,
for the $\nu_e$ (red), $\bar\nu_e$ (blue), $\nu_\mu$ (green),  and $\bar\nu_\mu$ (purple)
channels.}
\label{fig:Deviation}
\end{figure}

We also plot the deviation of the reconstructed neutrino energy, as a
fraction of the true neutrino energy, in Figure~\ref{fig:Deviation} for the $\nu_e$ (red), $\bar\nu_e$ (blue), $\nu_\mu$ (green)  and $\bar\nu_\mu$ (purple)
channels. We see from Figure~\ref{fig:Deviation} that the mean of the reconstructed neutrino energy distribution lies very close to the true neutrino energy.  This can be contrasted with the analysis of~\cite{GrantYang}, in which it was necessary to shift the
mean of the reconstructed energy distribution by a significant offset, which must be determined by either simulation or calibration.

Lastly, we determine the number of expected background events which pass all of the constraints, assuming atmospheric neutrino
fluxes given in Eq.~\ref{eq:HomestakeFlux}.
With an exposure of 380~kT~yr, the combined number of charged current events initiated by atmospheric
($\nu_e, \bar \nu_e$), ($\nu_\mu, \bar \nu_\mu$) with energy in the 1-24~GeV range, is
$\sim 1.6 \times 10^4$ ($3.3 \times 10^4$). The number of expected background CC-events passing the cuts and with
reconstructed energy lying within a
window of size given by the energy resolutions is shown in Table~\ref{tab:NumberOfBackgroundPassingCuts}.  Indeed, given this choice of cuts, we find $\lesssim 1$ background event in the ($\nu_e,\bar\nu_e$) channels
for any choice of mass in our mass range.

\begin{table}
\centering
\begin{tabular}{|c|c|c|c|c|}
  \hline
  $E$[GeV] & $N_{B,e}$ & $N_{B,\mu}$ & $\sigma_{S,e}$[GeV] & $\sigma_{S,\mu}$[GeV]\\
  \hline
4	&	1.095	&	3.723	&	0.521	&	0.763 \\
5	&	0.893	&	3.592	&	0.652	&	0.955	\\
6	&	0.659	&	3.457	&	0.784	&	1.149	\\
7	&	0.669	&	3.073	&	0.920	&	1.344	\\
8	&	0.571	&	2.911	&	1.069	&	1.547	\\
9	&	0.505	&	2.698	&	1.219	&	1.751	\\
10	&	0.465	&	2.474	&	1.369	&	1.974	\\
11	&	0.395	&	2.330	&	1.539	&	2.180	\\
12	&	0.374	&	2.126	&	1.712	&	2.402	\\
13	&	0.322	&	1.962	&	1.891	&	2.617	\\
14	&	0.330	&	1.844	&	2.068	&	2.843	\\
15	&	0.312	&	1.748	&	2.258	&	3.073	\\
  \hline
\end{tabular}
\caption{The expected number of background events which pass all the cuts and are reconstructed to be in an energy window of the size given by the energy resolution, centered at 1 GeV increments in the range 4-15 GeV.  The corresponding energy resolutions are given in the fourth and the fifth columns. $N_{B,e}$ indicates the expected number of (anti-)electron neutrino background events passing the cuts out of total $1.6 \times 10^4$ expected CC scattering events, and $N_{B,\mu}$ the expected number of (anti-)muon neutrino background events passing the cuts out of
total $3.3 \times 10^4$ expected CC scattering events.}
\label{tab:NumberOfBackgroundPassingCuts}
\end{table}

\section{Sensitivity to Dark Matter Annihilation}

Dark matter can be gravitationally captured in the Sun if it elastically scatters off solar nuclei, and in the process loses enough
kinetic energy to nuclear recoil to drop below the escape velocity~\cite{Gould:1987ir,Gould:1991hx}. The rate at which dark matter is captured thus depends on
the dark matter mass ($m_{\chi}$), and is proportional to the scattering cross section.  We focus on the case in which dark matter scattering
is spin-dependent (SD), as this is the case most difficult to probe with direct detection experiments, but which can be probed by searches for
dark matter scattering off hydrogen in the Sun.  The dark matter capture rate can then be expressed as
$\Gamma_C = [C_0^{\rm SD} (m_{\chi}) \times \sigma_{\rm SD}] (\frac{\rho_{\odot}}{0.3\gev/\cm^3}) (\frac{\bar v}{ 270~\km/\s})^{-1}$, where $\sigma_{\rm SD}$ is the dark matter-proton spin-dependent scattering cross section,
$\rho_{\odot}$ is the assumed average dark matter halo density at the solar radius and $\bar v$ is the dark matter velocity dispersion.
$C_0^{\rm SD} (m_{\chi})$ is a function which depends only on the dark matter mass, known solar physics, and standard assumptions about the dark matter
density and velocity distributions. It should be noted that the dependence on the dark matter velocity distribution and circular velocity of the Sun have only minor impacts on the capture rates in the Sun and effects are in general smaller than 20\%~\cite{Choi:2013eda,Danninger:2014xza}. Recent data supports lower dark matter velocity distributions~\cite{Necib:2018iwb}, which is expected to enhance solar dark matter signals while reducing events rate in direct detection experiments. For the relevant mass range, the $C_0^{\rm SD} (m_{\chi})$ can be found, for example, in~\cite{Gao:2011bq,Kumar:2012uh}.

We assume that dark matter annihilates to monoenergetic neutrinos via the process $\chi\chi \rightarrow \bar \nu \nu$.
This scenario provides a striking signal at neutrino detectors, which is useful for illustrating the utility of the
reconstruction algorithm which we have described.
But note that this process will not be realized in standard scenarios of supersymmetry (SUSY), because the process
$\chi \chi \rightarrow \bar \nu \nu$ is chirality/$p$-wave suppressed if the dark matter particle is its own anti-particle and if its interactions
respect minimal flavor violation (see, for example~\cite{Kumar:2013iva}).
But as there is still no experimental evidence for low-scale SUSY, one can easily realize
scenarios in which dark matter annihilates dominantly to monoenergetic neutrinos if dark matter is a
Dirac fermion ($\bar \chi \chi \rightarrow \bar \nu \nu$), or if flavor-violation is not minimal.

If dark matter is in equilibrium in the Sun, as it will be for the region of parameter space we consider~\cite{Kumar:2012uh}, then the dark matter annihilation rate is
related to the capture rate by $\Gamma_A = (1/2) \Gamma_C$.
The flux of (anti-)neutrinos arriving from the direction of the Sun due to dark matter annihilation is given by
\bea
{d\Phi_{\rm DM}^{\nu_\ell, \bar \nu_\ell} \over  dE} &=&\frac{B_{\nu_\ell \bar \nu_\ell}}{8\pi r_\oplus^2} C_0^{\rm SD} (m_{\chi}) \sigmaSD^p
\left[\left( {\rho_{\odot} \over 0.3\gev/\cm^3} \right) \left({\bar v \over 270~\km / \s} \right)^{-1}  \right] \delta (E_\nu - m_{\chi}),
\label{eq:DMFluxFull}
\eea
where $r_\oplus$ is the Earth-Sun distance
and $B_{\nu_\ell \bar \nu_\ell}$ is the branching fraction for dark matter annihilation process $\chi\chi \rightarrow \bar \nu_\ell \nu_\ell$, and we assume
that these branching fractions are flavor-independent (in this case, neutrino oscillations have no effect).  Note, we assume that $m_{\chi} \gtrsim 4\gev$, in
which case the evaporation of dark matter from the Sun is negligible~\cite{WIMPevaporation}.

\begin{figure}[ht]
\centering
\includegraphics[width=0.8\textwidth]{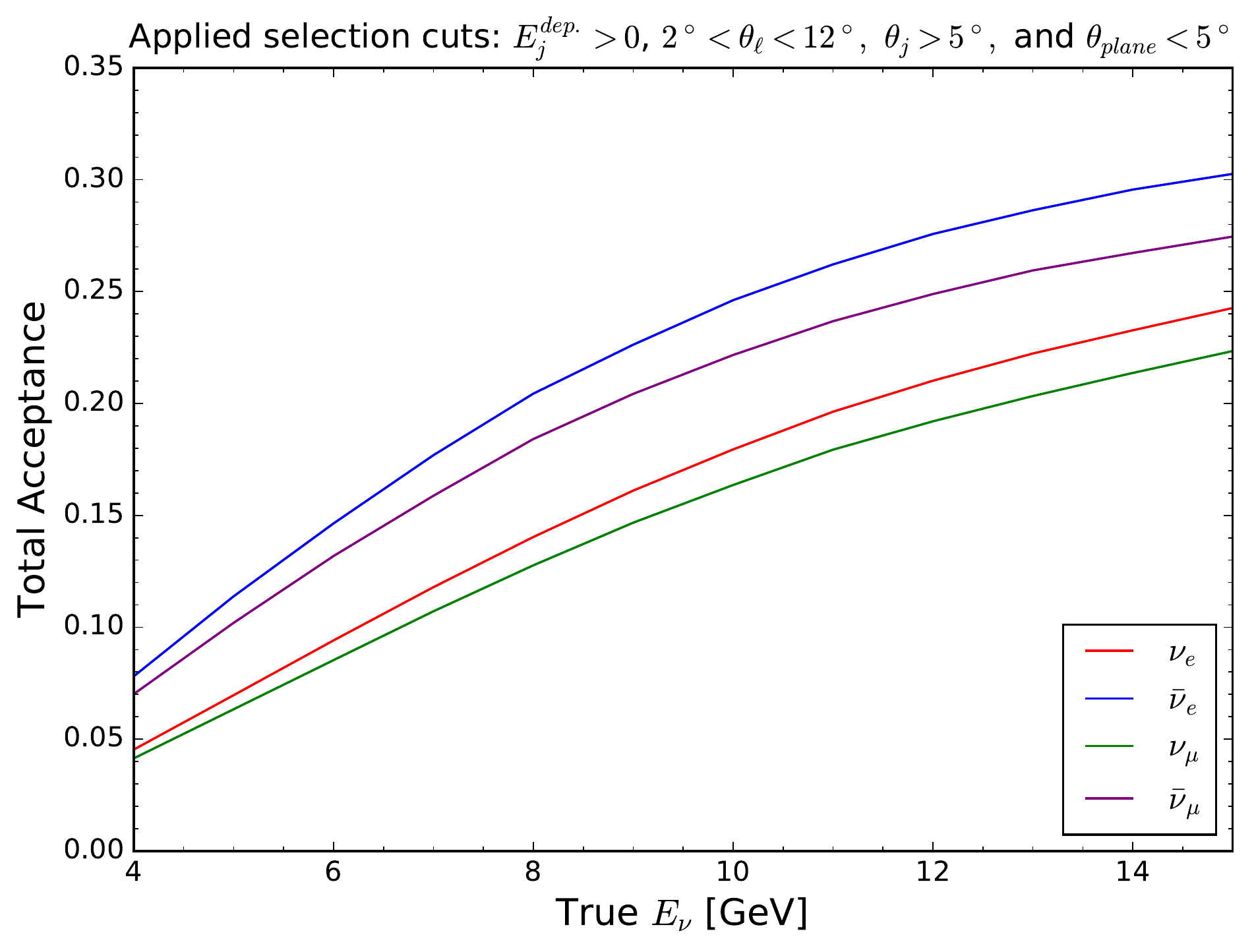}
\caption{Plot of the acceptance, $f_{\rm DM}^{\nu_\ell, \bar \nu_\ell}$, as a function of the true neutrino
energy, for the $\nu_e$ (red), $\bar \nu_e$ (blue), $\nu_\mu$ (green),  and $\bar \nu_\mu$ (purple)
channels.}
\label{fig:Acceptance}
\end{figure}

The number of expected events arising from dark matter annihilation in which the (anti-)neutrino energy is reconstructed within the energy
window $E_\nu \pm \Delta E/2$ is then given by
\bea
N^{\nu_\ell, \bar \nu_\ell}_{\rm DM} (E_\nu, \Delta E) &=&  f_{\rm DM}^{\nu_\ell, \bar \nu_\ell} (E_\nu, \Delta E)
\int_{E_\nu - \Delta E /2}^{E_\nu+ \Delta E/2} dE \,
{d \Phi_{\rm DM}^{\nu_\ell, \bar \nu_\ell} \over dE} \times  \left( T A^{\rm eff.}_{\nu, \bar \nu} \right) ,
\label{eq:NumSigExpected}
\eea
where
\bea
T A^{\rm eff.}_{\nu, \bar \nu} &=& \left( \frac{\rm exposure}{\kT ~\yr} \right) \left(1.9 \times 10^{40} \s  \right) \sigma_{\nu, \bar \nu N} ,
\eea
and $f_{\rm DM}^{\nu_\ell, \bar \nu_\ell}$ is the fraction of events with true neutrino energy $E_\nu$ which satisfy the angular cuts and have a
reconstructed energy lying within the window $E_\nu \pm \Delta E/2$.  We plot $f_{\rm DM}^{\nu_\ell, \bar \nu_\ell}$ for all four channels in
Figure~\ref{fig:Acceptance}.

For any choice of the parameters $(m_{\chi}, \sigma_{\rm SD})$, Eq.~\ref{eq:NumSigExpected} gives
the number of expected signal events in each of the four (anti-)neutrino channels.  Given our estimate for the number of background events, one
can then estimate the potential sensitivity of DUNE.  We will adopt the following criteria.  We assume that the number of events actually observed
in each channel is consistent with the expected number of background events.  A choice of the parameters   $(m_{\chi}, \sigma_{\rm SD})$ is then considered to be
excluded at 90\% CL if, under the assumption that this parameter choice is the true model,  the likelihood of producing the observed number of events or
fewer in each of the considered channels is 10\% or less.

\begin{figure}[ht]
\centering
\includegraphics[width=1\textwidth]{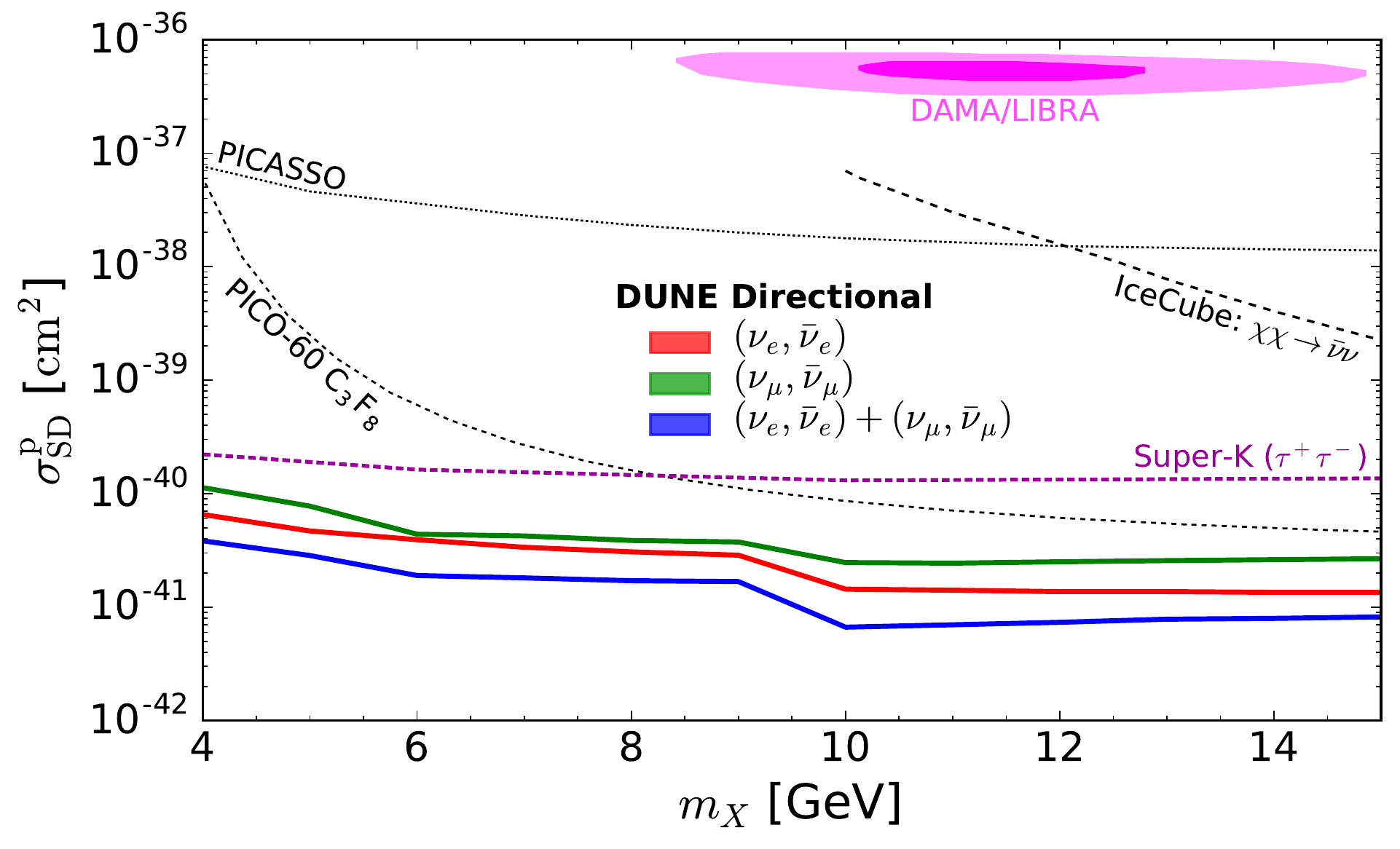}
\caption[DUNE]{The expected sensitivity of DUNE, assuming a 380~kT yr exposure is shown.  We consider the ($\nu_e,\bar\nu_e$) channels (red), the ($\nu_\mu,\bar\nu_\mu$) channels (green), and a combined analysis (blue).  For comparison we show the region consistent with the DAMA/LIBRA signal (at 90\%/3$\sigma$ CL)~\cite{Savage:2008er} and current indirect bounds from  IceCube~\cite{Aartsen:2016exj} and Super-Kamiokande~\cite{Choi:2015ara}. For Super-Kamiokande no official collaboration result on dark matter annihilation into neutrinos is available, therefore we show the $\tau^{+}\tau^{-}$ channel for comparison. Directed detection bounds from PICO-60~\cite{Amole:2017dex} and PICASSO~\cite{PICASSO2016} are shown. In the latter case we use a conservative choice of the intrinsic energy resolution and quenching factor.}
\label{fig:Sensitivity}
\end{figure}

We plot the resulting sensitivity of our analysis for 380~kT yr exposure in Figure~\ref{fig:Sensitivity}, assuming a branching fraction of $B_{\nu_\ell \bar \nu_\ell}=1/3$ with $\ell = e, \mu,$ or $\tau$. We consider the ($\nu_e,\bar\nu_e$) channels (red), the ($\nu_\mu,\bar\nu_\mu$) channels (green),
and a combined analysis of all four channels (blue).
Bounds from indirect searches from IceCube~\cite{Aartsen:2016exj} and Super-Kamiokande~\cite{Choi:2015ara}, and direct detection bounds from PICASSO~\cite{PICASSO2016} and PICO-60~\cite{Amole:2017dex} are shown for comparison together with the region consistent with the DAMA/LIBRA signal~\cite{Savage:2008er}. In our comparison, we show the latest published result from PICO-60, and note that there is a new preliminary limit~\cite{Amole:2019fdf}, with improved sensitivity in the 3-5 GeV mass range.
But we note that the Super-Kamiokande bound is for the
$\tau^+ \tau^-$ channel, as Super-Kamiokande has not performed an analysis for the monoenergetic neutrino channel; such an analysis would
be expected to have greater sensitivity.

Note that in our analysis ($\nu_e,\bar\nu_e$) channels are expected to provide a slightly greater sensitivity than the
($\nu_\mu,\bar\nu_\mu$) channels; this is driven largely by the fact that the ($\nu_e,\bar\nu_e$) are expected to produce contained showers whose energy can be measured with
greater precision than the ($\nu_\mu,\bar\nu_\mu$), whose tracks will often exit the detector.

It is interesting to compare these results to those of~\cite{Kumar:2015nja}, which considered the sensitivity of a $\sim 40$~kT~yr detector to
monoenergetic neutrinos arising from dark matter annihilation in the Sun, using the ($\nu_e,\bar\nu_e$) channels.
For $m_X \sim 10\gev$, the expected sensitivity found in the analysis of~\cite{Kumar:2015nja} was $\sim 10^{-40}~\cm^2$,
which is a factor $\sim 10$ worse than that found here, for a factor $\sim 10$ smaller exposure.  These results are consistent, since in both cases it was found
that the number of background events in the ($\nu_e,\bar\nu_e$) channels was negligible; in such a case, sensitivity scales linearly with exposure.  Interestingly, it was found in~\cite{Kumar:2015nja} that
one would expect $\sim 0.1$ background events passing the cuts over a 40~kT~yr exposure, assuming optimistically that the neutrino energy resolution of the detector was $\sim 3\%$.  Even assuming a
factor $\sim 10$ larger exposure and a factor $\sim 7$ worse energy resolution, we have still found $\lesssim 1$ expected background event in the ($\nu_e,\bar\nu_e$) channels.  The reduction in background
is largely due to the effect of requiring the jet direction to lie near the plane of the event, a technique not utilized in~\cite{Kumar:2015nja}.

\section{Conclusions}

We have presented a new procedure for reconstructing the energy of an incoming neutrino from the tracks produced
in an LArTPC detector after a charged-current interaction, assuming that the direction of the incoming neutrino is known.
This procedure differs from others which have been considered
by its use of the observed jet direction and kinematic constraints to reduce the uncertainty arising from jet mismeasurement.
Although these techniques are familiar in the context of high energy collider analyses, we have found that they can be
effectively applied to neutrino detectors.

In addition to providing an alternative and competitive method for reconstructing the neutrino energy, this method also provides
for a much improved method for rejecting background events arising from atmospheric neutrinos, whose direction is not correlated to the direction
of the source of interest.  The key to this improved background rejection is the use of the jet direction in conjunction with the charged lepton
direction.  It is well-known that, when an energetic neutrino produces a charged lepton through a CC-interaction, the lepton tends
to be produced within a cone pointing in the forward direction, and this feature has been used to reject background in past searches for
neutrinos arising from DM annihilation in the Sun.  However, the direction of the jet has not been used thus far, since it is difficult to determine
its direction in water Cherenkov  detectors, though it can be determined with good resolution in
an LArTPC detector.  Not only must the jet also be produced in the forward direction, but the plane spanned by the momenta of the lepton
and the jet must also include the Sun.  Inclusion of jet direction information thus reduces the phase space consistent with event topology
from a 3D cone to a 2D surface (the intersection of the cone with a plane), yielding a dramatic improvement in background rejection which
is limited only by the angular resolution of the detector.

Although we have applied this energy and topology reconstruction technique to the search for neutrinos arising from dark matter annihilation
in the Sun, the fundamental idea can be applied to any situation in which the neutrinos arrive from a known direction.  In particular, this
technique can also be used to reconstruct events in which the neutrino arrives from the Fermilab beamline.  However, for this case, one may choose
a different set of event selection cuts.  The cuts we have used here were chosen to minimize the atmospheric neutrino background to a dark
matter search.  Neutrinos arriving from the beamline have a larger flux and arrive at known times, making the atmospheric neutrino background
less problematic.  For that purpose, one may instead choose alternate selection cuts which improve signal acceptance or energy resolution.

We expect that the ability to reject background should improve at lower energies.  The reason is that,
at higher energies, the forward cone in which the lepton is produced by the CC-interaction is narrower, and once the angular
resolution of the detector is accounted for, it is more difficult to distinguish events for which the lepton and the jet do not span a plane
containing the source.  However, conversely, one would expect background rejection to be improved in scenarios in which the neutrino has a lower
energy.  If dark matter annihilates to light quarks, then the resulting light mesons ($\pi^+$, $K^+$) will tend to decay at rest in the Sun,
producing monoenergetic 30~MeV and 236~MeV neutrinos~\cite{Bernal:2012qh,Rott:2012qb,Rott:2015nma,Rott:2016mzs,Rott:2017weo}, providing interesting signals for DUNE and Hyper-K~\cite{Abe:2015zbg,Abe:2011ts}.  When a 236~MeV neutrino scatters off an
argon nucleus via a CC interaction, the produced charged lepton is emitted almost isotropically, but a large fraction of events contain a proton
which is preferentially ejected in the forward direction.  It has been shown that one can greatly reduce backgrounds by searching for events
with 236~MeV reconstructed energy in which the ejected proton points away from the source, but our results suggest that a much more dramatic
improvement could be obtained by requiring the plane spanned by the charged lepton and the proton to contain the source.  It would be interesting
to study this strategy quantitatively.

We have considered the scenario in which dark matter in the Sun annihilates to monoenergetic neutrinos ($\chi \chi \rightarrow \bar \nu \nu$),
because searches for monoenergetic neutrinos from the Sun can provide clear discovery potential through an unmistakable signal.
However, there is no known Standard Model process that could produce high-energy monoenergetic neutrinos from the Sun.
Neutrinos from the solar atmosphere provide a natural sensitivity floor to solar dark matter searches as has recently been pointed out in several publications~\cite{Arguelles:2017eao,Ng:2017aur,Edsjo:2017kjk,Masip:2017gvw}. A dark matter signal with a clear energy feature, such as a monoenergetic neutrino line, can however more easily  be discriminated from this background.

If neutrinos are not produced directly as final state of the dark matter annihilations, we expect that in almost all dark matter scenarios neutrinos are created through sequential decays of the annihilation products. In many cases this yields a dominant production channel leading to some feature in the neutrino energy spectrum. Secluded dark matter models (example $\chi\chi \rightarrow Y Y \rightarrow \nu \nu \nu \nu$) can produce box-shape spectra~\cite{Leane:2017vag,Pospelov:2007mp} and Standard Model particle decays result in a variety of spectra~\cite{Bringmann:2018lay,Cirelli:2010xx}. If the spectral feature is similar in size to the detector energy resolution, one can also reinterpret our results with appropriate scaling.  But for any neutrino energy spectrum produced by a source in a fixed direction, our results
indicate the extent to which the energy spectrum will be smeared by the energy reconstruction algorithm.

We also note that a recent paper~\cite{Friedland:2018vry} has suggested that dramatic improvements
in DUNE's expected energy resolution, relative to the results of~\cite{GrantYang}, can be made by
utilizing information about quenching, as well as more optimistic estimates of the DUNE particle identification
thresholds.  We emphasize that these techniques are entirely complementary to the ones we have described here.
If one could apply those techniques to an analysis of the type we have described one would expect a much more
accurate measurement of the jet direction (since more particles in the hadronic cascade would be identified,
and their energies would be measured more accurately), leading to an improvement in neutrino energy resolution
and background rejection.

\vskip .2in
\acknowledgments

We are grateful to Jelena Maricic, Yujing Sun and Tingjun Yang for useful discussions.
The work of J.~Kumar is supported in part by DOE grant DE-SC0010504.
C.~Rott acknowledges support from the National Research Foundation of Korea (NRF) for the Basic Science Research Program NRF-2017R1A2B2003666. The work of D. Yaylali is supported in part by DOE grant DE-FG02-13ER41976 (de-sc0009913).


\appendix
\section{Neutrino Cross sections computed with NuWro}
\label{app:cross}

The cross sections used for our analysis are given in Tables~\ref{tab:NeutrinoNucleonScatteringCrossSectionElectron},~\ref{tab:NeutrinoNucleonScatteringCrossSectionAntiElectron},~\ref{tab:NeutrinoNucleonScatteringCrossSectionMuon}, and~\ref{tab:NeutrinoNucleonScatteringCrossSectionAntiMuon}, which give the cross section tables computed with \verb+NuWro+ for electron, anti-electron, muon, and anti-muon neutrinos, respectively.

\begin{table}[ht!]
\centering
\begin{tabular}{|c|c|c|c|c|c|c|}
  \hline
E[GeV]	&	QEL-CC	&	RES-CC	&	DIS-CC	&	COH-CC	&	MEC-CC	&	Total[$\cm^2$]	\\
\hline
1	&	5.57E-39	&	3.65E-39	&	2.04E-41	&	7.84E-41	&	1.11E-39	&	1.04E-38	\\
2	&	5.78E-39	&	6.91E-39	&	2.96E-39	&	1.70E-40	&	9.46E-40	&	1.68E-38	\\
3	&	5.64E-39	&	7.69E-39	&	9.00E-39	&	2.45E-40	&	8.92E-40	&	2.35E-38	\\
4	&	5.52E-39	&	7.91E-39	&	1.57E-38	&	3.14E-40	&	8.62E-40	&	3.03E-38	\\
5	&	5.44E-39	&	7.95E-39	&	2.26E-38	&	3.77E-40	&	8.43E-40	&	3.72E-38	\\
6	&	5.41E-39	&	7.96E-39	&	2.95E-38	&	4.36E-40	&	8.29E-40	&	4.41E-38	\\
7	&	5.35E-39	&	7.97E-39	&	3.64E-38	&	4.92E-40	&	8.23E-40	&	5.11E-38	\\
8	&	5.32E-39	&	7.96E-39	&	4.33E-38	&	5.44E-40	&	8.21E-40	&	5.80E-38	\\
9	&	5.30E-39	&	7.96E-39	&	5.03E-38	&	5.94E-40	&	8.14E-40	&	6.50E-38	\\
10	&	5.30E-39	&	7.97E-39	&	5.73E-38	&	6.40E-40	&	8.08E-40	&	7.20E-38	\\
11	&	5.29E-39	&	7.96E-39	&	6.42E-38	&	6.85E-40	&	8.14E-40	&	7.89E-38	\\
12	&	5.26E-39	&	7.96E-39	&	7.12E-38	&	7.27E-40	&	8.08E-40	&	8.59E-38	\\
13	&	5.22E-39	&	7.96E-39	&	7.82E-38	&	7.68E-40	&	8.04E-40	&	9.29E-38	\\
14	&	5.24E-39	&	7.97E-39	&	8.51E-38	&	8.07E-40	&	8.00E-40	&	9.99E-38	\\
15	&	5.26E-39	&	7.96E-39	&	9.21E-38	&	8.45E-40	&	8.00E-40	&	1.07E-37	\\
16	&	5.19E-39	&	7.99E-39	&	9.93E-38	&	8.80E-40	&	8.02E-40	&	1.14E-37	\\
17	&	5.17E-39	&	8.00E-39	&	1.06E-37	&	9.13E-40	&	7.97E-40	&	1.21E-37	\\
18	&	5.18E-39	&	7.96E-39	&	1.13E-37	&	9.45E-40	&	8.21E-40	&	1.28E-37	\\
19	&	5.31E-39	&	7.92E-39	&	1.20E-37	&	9.84E-40	&	7.94E-40	&	1.35E-37	\\
20	&	5.16E-39	&	7.95E-39	&	1.27E-37	&	1.01E-39	&	7.77E-40	&	1.42E-37	\\
21	&	5.15E-39	&	7.89E-39	&	1.33E-37	&	1.05E-39	&	8.14E-40	&	1.48E-37	\\
22	&	5.06E-39	&	7.92E-39	&	1.40E-37	&	1.07E-39	&	7.79E-40	&	1.55E-37	\\
23	&	5.18E-39	&	7.84E-39	&	1.48E-37	&	1.10E-39	&	7.97E-40	&	1.63E-37	\\
24	&	5.17E-39	&	7.86E-39	&	1.55E-37	&	1.13E-39	&	7.86E-40	&	1.70E-37	\\
  \hline
\end{tabular}
\caption[Neutrino-Nucleon CC Scattering Cross-section for $\nu_e$]{$\sigma_{\nu N}$ (in $\cm^2$) for the various subprocesses arising from $\nu_e + {}^{40}{\rm Ar}$ scattering for
incoming neutrino energy $1-24$ GeV. The subprocesses are labelled as QEL (quasi-elastic), RES (resonant production), DIS (deep inelastic), COH (coherent), and MEC (meson exchange current).}
\label{tab:NeutrinoNucleonScatteringCrossSectionElectron}
\end{table}

\begin{table}[ht!]
\centering
\begin{tabular}{|c|c|c|c|c|c|c|}
  \hline
E[GeV]	&	QEL-CC	&	RES-CC	&	DIS-CC	&	COH-CC	&	MEC-CC	&	Total[$\cm^2$]	\\
\hline
1	&	1.40E-39	&	8.94E-40	&	4.64E-42	&	7.87E-41	&	2.61E-40	&	2.64E-39	\\
2	&	2.31E-39	&	2.61E-39	&	5.68E-40	&	1.70E-40	&	4.68E-40	&	6.13E-39	\\
3	&	2.81E-39	&	3.77E-39	&	1.95E-39	&	2.46E-40	&	5.63E-40	&	9.34E-39	\\
4	&	3.10E-39	&	4.52E-39	&	3.85E-39	&	3.14E-40	&	6.10E-40	&	1.24E-38	\\
5	&	3.28E-39	&	5.05E-39	&	6.08E-39	&	3.77E-40	&	6.43E-40	&	1.54E-38	\\
6	&	3.42E-39	&	5.45E-39	&	8.51E-39	&	4.36E-40	&	6.61E-40	&	1.85E-38	\\
7	&	3.51E-39	&	5.73E-39	&	1.11E-38	&	4.91E-40	&	6.82E-40	&	2.15E-38	\\
8	&	3.61E-39	&	5.96E-39	&	1.37E-38	&	5.43E-40	&	6.88E-40	&	2.45E-38	\\
9	&	3.65E-39	&	6.18E-39	&	1.64E-38	&	5.93E-40	&	7.03E-40	&	2.76E-38	\\
10	&	3.70E-39	&	6.33E-39	&	1.92E-38	&	6.40E-40	&	7.10E-40	&	3.06E-38	\\
11	&	3.77E-39	&	6.46E-39	&	2.21E-38	&	6.84E-40	&	7.14E-40	&	3.37E-38	\\
12	&	3.77E-39	&	6.54E-39	&	2.49E-38	&	7.27E-40	&	7.20E-40	&	3.67E-38	\\
13	&	3.84E-39	&	6.66E-39	&	2.78E-38	&	7.67E-40	&	7.23E-40	&	3.98E-38	\\
14	&	3.84E-39	&	6.73E-39	&	3.07E-38	&	8.07E-40	&	7.32E-40	&	4.28E-38	\\
15	&	3.86E-39	&	6.80E-39	&	3.36E-38	&	8.44E-40	&	7.36E-40	&	4.59E-38	\\
16	&	3.85E-39	&	6.87E-39	&	3.66E-38	&	8.82E-40	&	7.28E-40	&	4.90E-38	\\
17	&	4.03E-39	&	6.94E-39	&	3.95E-38	&	9.18E-40	&	7.21E-40	&	5.21E-38	\\
18	&	3.98E-39	&	6.98E-39	&	4.25E-38	&	9.48E-40	&	7.70E-40	&	5.51E-38	\\
19	&	3.90E-39	&	7.03E-39	&	4.58E-38	&	9.81E-40	&	7.41E-40	&	5.84E-38	\\
20	&	3.89E-39	&	7.08E-39	&	4.84E-38	&	1.01E-39	&	7.40E-40	&	6.11E-38	\\
21	&	3.89E-39	&	7.06E-39	&	5.12E-38	&	1.04E-39	&	7.59E-40	&	6.39E-38	\\
22	&	3.96E-39	&	7.17E-39	&	5.42E-38	&	1.07E-39	&	7.57E-40	&	6.71E-38	\\
23	&	3.98E-39	&	7.17E-39	&	5.74E-38	&	1.11E-39	&	7.48E-40	&	7.04E-38	\\
24	&	3.98E-39	&	7.18E-39	&	6.04E-38	&	1.13E-39	&	7.45E-40	&	7.34E-38	\\
\hline
\end{tabular}
\caption[Neutrino-Nucleon CC Scattering Cross-section for ($\bar\nu_e$)]{$\sigma_{\bar \nu N}$ (in cm$^2$) for the various subprocesses arising from $\bar\nu_e + {}^{40}{\rm Ar}$ scattering for
incoming neutrino energy $1-24$ GeV. The subprocesses are labelled as QEL (quasi-elastic), RES (resonant production), DIS (deep inelastic), COH (coherent), and MEC (meson exchange current).}
\label{tab:NeutrinoNucleonScatteringCrossSectionAntiElectron}
\end{table}

\begin{table}[ht!]
\centering
\begin{tabular}{|c|c|c|c|c|c|c|}
  \hline
E[GeV]	&	QEL-CC	&	RES-CC	&	DIS-CC	&	COH-CC	&	MEC-CC	&	Total[$\cm^2$]	\\
\hline
1	&	5.47E-39	&	3.42E-39	&	6.40E-42	&	6.12E-41	&	1.10E-39	&	1.01E-38	\\
2	&	5.76E-39	&	6.79E-39	&	2.75E-39	&	1.51E-40	&	9.47E-40	&	1.64E-38	\\
3	&	5.60E-39	&	7.65E-39	&	8.70E-39	&	2.27E-40	&	8.90E-40	&	2.31E-38	\\
4	&	5.52E-39	&	7.85E-39	&	1.53E-38	&	2.95E-40	&	8.61E-40	&	2.99E-38	\\
5	&	5.45E-39	&	7.92E-39	&	2.22E-38	&	3.58E-40	&	8.41E-40	&	3.67E-38	\\
6	&	5.39E-39	&	7.95E-39	&	2.91E-38	&	4.16E-40	&	8.31E-40	&	4.37E-38	\\
7	&	5.34E-39	&	7.96E-39	&	3.60E-38	&	4.71E-40	&	8.28E-40	&	5.06E-38	\\
8	&	5.31E-39	&	7.94E-39	&	4.29E-38	&	5.24E-40	&	8.22E-40	&	5.75E-38	\\
9	&	5.29E-39	&	7.94E-39	&	4.98E-38	&	5.72E-40	&	8.16E-40	&	6.45E-38	\\
10	&	5.29E-39	&	7.95E-39	&	5.68E-38	&	6.19E-40	&	8.10E-40	&	7.15E-38	\\
11	&	5.25E-39	&	7.96E-39	&	6.38E-38	&	6.63E-40	&	8.05E-40	&	7.85E-38	\\
12	&	5.26E-39	&	7.95E-39	&	7.07E-38	&	7.06E-40	&	8.04E-40	&	8.54E-38	\\
13	&	5.24E-39	&	7.93E-39	&	7.77E-38	&	7.46E-40	&	8.02E-40	&	9.24E-38	\\
14	&	5.25E-39	&	7.94E-39	&	8.46E-38	&	7.85E-40	&	8.04E-40	&	9.94E-38	\\
15	&	5.22E-39	&	7.91E-39	&	9.15E-38	&	8.23E-40	&	7.99E-40	&	1.06E-37	\\
16	&	5.27E-39	&	7.95E-39	&	9.84E-38	&	8.57E-40	&	8.14E-40	&	1.13E-37	\\
17	&	5.24E-39	&	7.93E-39	&	1.05E-37	&	8.93E-40	&	7.97E-40	&	1.20E-37	\\
18	&	5.02E-39	&	7.97E-39	&	1.12E-37	&	9.28E-40	&	7.95E-40	&	1.27E-37	\\
19	&	5.20E-39	&	7.95E-39	&	1.19E-37	&	9.55E-40	&	7.85E-40	&	1.33E-37	\\
20	&	5.23E-39	&	7.90E-39	&	1.26E-37	&	9.88E-40	&	7.74E-40	&	1.41E-37	\\
21	&	5.31E-39	&	7.96E-39	&	1.34E-37	&	1.02E-39	&	7.70E-40	&	1.49E-37	\\
22	&	5.19E-39	&	7.92E-39	&	1.40E-37	&	1.05E-39	&	7.97E-40	&	1.55E-37	\\
23	&	5.18E-39	&	7.93E-39	&	1.48E-37	&	1.08E-39	&	8.01E-40	&	1.63E-37	\\
24	&	5.10E-39	&	7.93E-39	&	1.54E-37	&	1.11E-39	&	7.89E-40	&	1.69E-37	\\
  \hline
\end{tabular}
\caption[Neutrino-Nucleon CC Scattering Cross-section for $\nu_\mu$]{$\sigma_{\nu N}$ (in $\cm^2$) for the various subprocesses arising from $\nu_\mu + {}^{40}{\rm Ar}$ scattering for incoming neutrino
energy $1-24$ GeV. The subprocesses are labelled as QEL (quasi-elastic), RES (resonant production), DIS (deep inelastic), COH (coherent), and MEC (meson exchange current).}
\label{tab:NeutrinoNucleonScatteringCrossSectionMuon}
\end{table}

\begin{table}[ht!]
\centering
\begin{tabular}{|c|c|c|c|c|c|c|}
  \hline
E[GeV]	&	QEL-CC	&	RES-CC	&	DIS-CC	&	COH-CC	&	MEC-CC	&	Total[$\cm^2$]	\\
\hline
1	&	1.39E-39	&	8.18E-40	&	1.15E-42	&	6.12E-41	&	2.58E-40	&	2.53E-39	\\
2	&	2.32E-39	&	2.57E-39	&	5.03E-40	&	1.51E-40	&	4.68E-40	&	6.01E-39	\\
3	&	2.79E-39	&	3.72E-39	&	1.85E-39	&	2.26E-40	&	5.59E-40	&	9.13E-39	\\
4	&	3.08E-39	&	4.47E-39	&	3.72E-39	&	2.94E-40	&	6.10E-40	&	1.22E-38	\\
5	&	3.28E-39	&	5.01E-39	&	5.92E-39	&	3.58E-40	&	6.41E-40	&	1.52E-38	\\
6	&	3.40E-39	&	5.41E-39	&	8.33E-39	&	4.16E-40	&	6.65E-40	&	1.82E-38	\\
7	&	3.49E-39	&	5.74E-39	&	1.09E-38	&	4.72E-40	&	6.80E-40	&	2.13E-38	\\
8	&	3.61E-39	&	5.95E-39	&	1.35E-38	&	5.24E-40	&	6.92E-40	&	2.43E-38	\\
9	&	3.63E-39	&	6.16E-39	&	1.62E-38	&	5.72E-40	&	6.96E-40	&	2.73E-38	\\
10	&	3.71E-39	&	6.29E-39	&	1.90E-38	&	6.19E-40	&	7.06E-40	&	3.03E-38	\\
11	&	3.76E-39	&	6.41E-39	&	2.18E-38	&	6.63E-40	&	7.14E-40	&	3.33E-38	\\
12	&	3.79E-39	&	6.56E-39	&	2.46E-38	&	7.06E-40	&	7.24E-40	&	3.64E-38	\\
13	&	3.83E-39	&	6.65E-39	&	2.76E-38	&	7.47E-40	&	7.27E-40	&	3.95E-38	\\
14	&	3.82E-39	&	6.74E-39	&	3.04E-38	&	7.85E-40	&	7.29E-40	&	4.25E-38	\\
15	&	3.87E-39	&	6.79E-39	&	3.33E-38	&	8.22E-40	&	7.31E-40	&	4.55E-38	\\
16	&	3.84E-39	&	6.86E-39	&	3.61E-38	&	8.58E-40	&	7.32E-40	&	4.84E-38	\\
17	&	3.83E-39	&	6.89E-39	&	3.93E-38	&	8.93E-40	&	7.46E-40	&	5.17E-38	\\
18	&	3.96E-39	&	6.97E-39	&	4.21E-38	&	9.26E-40	&	7.38E-40	&	5.47E-38	\\
19	&	3.97E-39	&	7.00E-39	&	4.52E-38	&	9.60E-40	&	7.46E-40	&	5.79E-38	\\
20	&	3.89E-39	&	7.06E-39	&	4.82E-38	&	9.92E-40	&	7.32E-40	&	6.09E-38	\\
21	&	3.96E-39	&	7.13E-39	&	5.13E-38	&	1.02E-39	&	7.35E-40	&	6.42E-38	\\
22	&	3.95E-39	&	7.14E-39	&	5.40E-38	&	1.05E-39	&	7.56E-40	&	6.69E-38	\\
23	&	3.87E-39	&	7.18E-39	&	5.71E-38	&	1.08E-39	&	7.47E-40	&	6.99E-38	\\
24	&	3.86E-39	&	7.12E-39	&	6.04E-38	&	1.10E-39	&	7.34E-40	&	7.32E-38	\\
\hline
\end{tabular}
\caption[Neutrino-Nucleon CC Scattering Cross-section for ($\bar\nu_\mu$)]{$\sigma_{\bar \nu N}$ (in cm$^2$) for the various subprocesses arising from $\bar\nu_\mu + {}^{40}{\rm Ar}$ scattering for incoming neutrino energy $1-24$ GeV. The subprocesses are labelled as QEL (quasi-elastic), RES (resonant production), DIS (deep inelastic), COH (coherent), and MEC (meson exchange current).}
\label{tab:NeutrinoNucleonScatteringCrossSectionAntiMuon}
\end{table}



\end{document}